\documentclass[twocolumn,           
               showpacs,            
               preprintnumbers,     
               aps,                 
               prd,          	    
letterpaper,           
               superscriptaddress,     
               nofootinbib,        
               tightenlines,        
               floats,floatfix      
               ]{revtex4-1}
\usepackage{graphicx}  
\usepackage{dcolumn}   
\usepackage{bm}       
\usepackage{amsmath,amssymb}
\usepackage{float}
\usepackage{makecell}
\usepackage[thinlines]{easytable}
\usepackage{array,booktabs}
\usepackage{longtable}
\usepackage{cancel}
\usepackage{array,makecell}

\begin{document} 

\title{New general parametrization of quintessence fields and its observational constraints}
\author{Nandan Roy}  
 \email{nandan@fisica.ugto.mx}
 \affiliation{%
Departamento de F\'isica, DCI, Campus Le\'on, Universidad de
Guanajuato, 37150, Le\'on, Guanajuato, M\'exico.}

\author{Alma X. Gonzalez-Morales}%
 \email{alma.gonzalez@fisica.ugto.mx}
 \affiliation{Consejo Nacional de Ciencia y Tecnolog\'ia,
Av. Insurgentes Sur 1582. Colonia Cr\'edito Constructor, Del. Benito Juárez C.P. 03940, M\'exico D.F. M\'exico}
\affiliation{%
Departamento de F\'isica, DCI, Campus Le\'on, Universidad de
Guanajuato, 37150, Le\'on, Guanajuato, M\'exico.}

\author{L. Arturo Ure\~{n}a-L\'{o}pez} 
 \email{lurena@ugto.mx}
\affiliation{%
Departamento de F\'isica, DCI, Campus Le\'on, Universidad de
Guanajuato, 37150, Le\'on, Guanajuato, M\'exico.}

\date{\today}

\begin{abstract}
We present a new parameterization of quintessence potentials for dark energy based directly upon the dynamical properties of the equations of motion. Such parameterization arises naturally once the equations of motion are written as a dynamical system in terms of properly defined polar variables. We have identified two different classes of parameters, and we dubbed them as dynamical and passive parameters. The dynamical parameters appear explicitly in the equations of motion, but the passive parameters play just a secondary role in their solutions. The new approach is applied to the so-called thawing potentials and it is argued that only three dynamical parameters are sufficient to capture the evolution of the quintessence fields at late times. This work reconfirms the arbitrariness of the quintessence potentials as the recent observational data fail to constrain the dynamical parameters.

\end{abstract}

\pacs{98.80.-k; 95.36.+x}

\keywords{cosmology, dynamical systems, phase space }

\maketitle

\section{Introduction \label{sec:introduction-}}

One of the most famous unsolved mysteries in modern Cosmology is the accelerated expansion of the universe, an observation that has been widely confirmed ever since its discovery in 1998\cite{perlmutter1999astrophys, riess1998observational,spergel2007three,tegmark2004cosmological,seljak2005cosmological,ade2016planck}. The accelerated expansion is commonly attributed to a mysterious matter component generically dubbed as dark energy (DE). The most accepted DE model is the cosmological constant\cite{Weinberg:1988cp,Peebles:2002gy,padmanabhan2003cosmological}, which is in fact part of the so-called standard model of Cosmology\cite{ade2016planck}. From this point of view, a cosmological constant represents a constant vacuum energy which can explain the accelerated expansion very well, but its existence is problematic from the theoretical point of view\cite{Weinberg:1988cp,Sahni:1999qe,Sahni:1999gb,Bianchi:2010uw}.

It seems then more natural to consider dynamical models where the DE component could be explained by extra fields in the matter budget or by modifications and/or extensions to our current understanding of the gravitational field\cite{copeland2006dynamics}. The latter possibility has been just recently weakened by the detection of gravitational waves produced during the collision of binary system of neutron stars, mainly because of the exquisite measurement that confirms that gravitational waves propagate at the speed of light\cite{Ezquiaga:2017ekz,Sakstein:2017xjx,Creminelli:2017sry,lombriser2016breaking,lombriser2017challenges}. Among the still surviving dynamical models of DE we find in particular those of quintessence scalar fields, which have been present in the literature for almost three decades \cite{ratra1988cosmological,tsujikawa2013quintessence,bamba2012dark}. In a quintessence model, a scalar field is minimally coupled to gravity and a potential supply the required negative pressure to drive the accelerated expansion of the universe. 

A wide range of quintessence potentials has been proposed in the literature\cite{Copeland:1997et,caldwell1998cosmological,zlatev1999quintessence,de2000cosmological,ng2001applications,corasaniti2003model,linder2006paths} but none of these have a confirmation from the observational point of view. Depending on the evolution of the equation of state parameter of the scalar field quintessence scalar field models are crudely classified into two classes \cite{caldwell2005limits,steinhardt1999cosmological,scherrer2008thawing,chiba2009slow} (i) thawing models and (ii) freezing models. For thawing models, the potential becomes shallow at late times and the field gradually slows down. For freezing models, during the early cosmological time, the field is almost frozen due to the presence of Hubble friction and the scalar field starts to slowly roll-down the potential as the field mass becomes lower than the Hubble expansion rate. For a more detailed discussion of the quintessence dynamics we refer to\cite{sahni20045, martin2008quintessence,linder2006paths,Bahamonde:2017ize}.

In this work we propose a general method to study the evolution of quintessence scalar field models with a general form of the scalar field potential. Using a suitable variable transformation, the equations of motion are written as a set of autonomous equations, which directly suggests a general parametrization of the quintessence potentials without having to know their precise form. Such a dynamical systems analysis of DE models is already popular in the study of cosmology, for examples and references see\cite{wainwright2005dynamical,coley2013dynamical,Garcia-Salcedo:2015ora,Bahamonde:2017ize}, but so far they have mostly used the original change of variables firstly introduced in Ref.~\cite{Copeland:1997et}.  The particular, polar form of the transformation into a dynamical system which we use in this work was first proposed for dark matter models and the inflationary scenario in~\citep{Urena-Lopez:2015gur,Urena-Lopez:2015odd}, but see also~\cite{reyes2010attractor,Roy:2013wqa} for other related works. As mentioned above, we shall show that there is a general parametrization from which almost all the popular quintessence potentials can be derived. The new parameters are the responsible for the dynamical behavior of the quintessence models, and we use this property to put observational constraints on their values and from this infer the functional form of the potentials that seem to be preferred by the data.

A summary of the paper is as follows. In Sec.~\ref{sec:math-backgr-}, we setup the mathematical background of the system. This section includes the formation of the autonomous system, its polar transformation, and a description of the general parametrization of the quintessence potentials. In addition, we provide a generic method to infer the potentials from the reverse integration of the given parametrization. In Sec.~\ref{sec:gener-solut-equat} we find approximate solutions to the equations of motion in their polar form to follow the evolution of the quintessence variables from the radiation dominated era up to the present time. As a result, we obtain analytical expressions that link initial values of the variables with present quantities of physical interest that can be used reliably in numerical solutions. Section~\ref{sec:numer-analys-results} is devoted to the numerical analysis of the quintessence solutions and their implementation in an amended version of the Boltzmann code CLASS. We also propose a new parametrization of the DE equation of state that suits well the behavior of the quintessence models, we study this by making a comparison with full numerical simulations of the equations of motion. The comparison with diverse cosmological observations is presented in Sec.~\ref{sec:numer-analys-results} by means of a full Bayesian analysis, in order to put constraints on the dynamical parameters of the quintessence models. Finally, we present a summary of our results and an outlook for future research in Sec.~\ref{sec:conclusions}.

\section{Mathematical background \label{sec:math-backgr-}}
We consider a flat Friedman-Lema\^itre-Robertson-Walker universe which
is dominated by the standard matter fluids plus a quintessence scalar
field. We also consider that all the component of the Universe are
barotropic in nature, i.e. the pressure $p_j$ and the density $\rho_j$
are related one to each other by the expression $p_j = w_j \rho_j $,
where $w_j$ are the corresponding (constant) equation of state (EOS)
parameter of each component. The Einstein field equations, the
continuity equation for each matter fluid, and the (wave) Klein-Gordon  
equation of scalar field can be written, respectively, as
\begin{subequations}
\label{eq:1}
  \begin{eqnarray}
    H^2 &=& \frac{\kappa^2}{3} \left( \sum_j \rho_j +
      \rho_\phi \right) \, , \label{eq:1a} \\
    \dot{H} &=& - \frac{\kappa^2}{2} \left[ \sum_j (\rho_j +
      p_j ) + (\rho_\phi + p_\phi) \right] \, , \label{eq:1b} \\
    \dot{\rho}_j &=& - 3 H (\rho_j + p_j ) \,
    , \label{eq:1c} \\
    \ddot{\phi} &=& -3 H \dot{\phi} - \frac{dV(\phi)}{d \phi} \, , \label{eq:1d}
  \end{eqnarray}
\end{subequations}
where $\kappa^2 = 8 \pi G$, $H \equiv \dot{a}/a$ is the Hubble parameter and $a$ the scale factor of the Universe, $V(\phi)$ is the scalar potential and a dot means derivative with respect to
cosmic time. The scalar field energy density $\rho_{\phi}$ and pressure $p_{\phi}$ are expressed in terms of the field variables, respectively, as
\begin{equation}
  \rho_\phi = \frac{1}{2} \dot{\phi}^2  + V ( \phi) \, ,
  \quad p_\phi = \frac{1}{2} \dot{\phi}^2 -  V( \phi) \, . \label{eq:17}
\end{equation}
In contrast to the corresponding quantities of the barotropic perfect fluids, the quintessence density and pressure cannot be handled independently and one necessarily requires to find separate solutions for $\phi$ and $\dot{\phi}$ from Eq.~\eqref{eq:1d}. In the sections below we will present new variables that help to solve Eqs.~\eqref{eq:1} more easily.

\subsection{Dynamical system approach \label{sec:dynam-syst-appr}}
To write Eq.(1d) as a set of autonomous
equations, we introduce a new set of dimensionless
variables\cite{Copeland:1997et,Urena-Lopez:2015odd,Roy:2014yta}
\begin{subequations}
  \label{eq:3}
  \begin{eqnarray}
    x &\equiv& \frac{\kappa \dot{\phi}}{\sqrt{6} H} \, , \quad y
               \equiv \frac{\kappa V^{1/2}}{\sqrt{3} H} \, , \label{eq:3a}
    \\
    y_1 &\equiv& - 2\sqrt{2} \frac{\partial_{\phi} V^{1/2}}{H} \, , \quad
                 y_2 \equiv - 4\sqrt{3} \frac{\partial^2_\phi
                 V^{1/2}}{\kappa H} \, . \label{eq:3b}             
  \end{eqnarray}
\end{subequations} 
As a result, Eq.(1d) is transformed into
\begin{subequations}
  \label{eq:4}
  \begin{eqnarray}
    x^\prime &=& - \frac{3}{2} \left(1 - w_{tot} \right) x + \frac{1}{2} y y_1 \,
                 , \label{eq:4a} \\
    y^\prime &=&  \frac{3}{2} \left(1 + w_{tot} \right) y - \frac{1}{2} x y_1 \,
                 , \label{eq:4b} \\
    y^\prime_1 &=&  \frac{3}{2} \left(1 + w_{tot} \right) y_1  + x
                   y_2 \, , \label{eq:4c}
  \end{eqnarray}
\end{subequations}
where now a prime is the derivative with respect to the number of
e-foldings, $N \equiv \ln(a/a_{i})$ and $a_i$ is the initial scale
factor of the universe. In writing Eqs.~\eqref{eq:4} we have used
the Friedmann constraint~\eqref{eq:1a} in the form $\Omega_r + \Omega_m + \Omega_\phi = 1$, where the density parameters of the different matter fields are defined in the standard way as $\Omega_j = \kappa^2 \rho_j/(3H^2)$. In addition, the total EoS is given by
\begin{equation} \label{eq.w}
w_{tot} \equiv \frac{p_{tot}}{\rho_{tot}} = \frac{1}{3} \Omega_r + x^2
- y^2 \, .
\end{equation}

Equations~\eqref{eq:4} have been thoroughly used in the literature to
study the properties of quintessence potentials, see for instance
Refs.\cite{Copeland:1997et,Bahamonde:2017ize} and references therein. Their main
advantage is the possibility to consider a compact phase space for the
variables $x$ and $y$, so that all trajectories and critical points of
interest can be studied without the intrinsic difficulties in the
standard variables $(\phi,\dot{\phi})$. One main limitation of this approach is that the form of the new potential variables $y_1$ and $y_2$ have to be calculated individually for each quintessence potential, and in this respect they do not offer a clear advantage over the direct solution of the KG equation~\eqref{eq:1d}, and the solution of the latter still is the popular approach in the numerical studies of quintessence models. 

\subsection{Polar form of the equations of motion \label{sec:polar-form-equations}}
We now introduce the following polar transformations of the variables: $x = \Omega_{\phi} ^{1/2} \sin(\theta/2)$ and $y = \Omega_{\phi}^{1/2}
\cos(\theta/2)$, where $\Omega_{\phi} = \kappa^2 \rho_{\phi}/3 H^2$ is
the density parameter associated to the quintessence field and $\theta$ represents an angular degree of freedom. The system
of equations~\eqref{eq:4}, after simple manipulations, reduces to
\begin{subequations}
\label{eq:10}
  \begin{eqnarray}
    \theta^{\prime} &=& - 3 \sin \theta + y_1 \, , \label{eq:10a} \\
    y_1^{\prime} &=& \frac{3}{2} \left( 1  + w_{tot} \right) y_1
                     + \Omega_{\phi} ^{1/2} \sin(\theta /2) y_2  \,
                     , \label{eq:10b} \\ 
    \Omega_{\phi} ^{\prime} &=& 3 (w_{tot} - w_{\phi})
                                \Omega_{\phi} \, . \label{eq:10c}       
  \end{eqnarray}
\end{subequations}
The EoS of the scalar field in terms of the polar variable is
$w_\phi = p_\phi/\rho_\phi = (x^2 - y^2)/(x^2 + y^2) = - \cos \theta$, which tells us of the direct relation between the two variables. Likewise, the ratio
of kinetic and potential energies is given by $\tan^2 \theta = (1/2) \dot{\phi}^2/V(\phi) = x^2/y^2$. Equations~\eqref{eq:10a} and~\eqref{eq:10c} are the same for any
kind of potential, and it is only Eq.~\eqref{eq:10b} that changes for
different cases because of the presence of $y_2$.

\subsection{General form of the quintessence potentials \label{sec:gener-form-quint}}
To find a solution of Eqs.~(\ref{eq:10}) one needs to close the system of
equations, and this can be done whenever the second potential variable
$y_2$ can be written in terms of the variables $\theta$, $y_1$ and $\Omega_\phi$. This is equivalent, in the standard approach, to the fixing of the scalar field potential. Then, one possibility is to choose the functional form of the potential $V(\phi)$ and to derive from it the form of $y_2$
following the prescriptions in Eqs.~\eqref{eq:3}. In Table~\ref{tab:1}
we give a list of thawing and freezing quintessence potentials that are very familiar in the current literature and their corresponding closed form in terms of $y_2$. For those potentials, the dynamical system~(\ref{eq:10}) becomes an autonomous one upon which we can use the known mathematical tools of such systems~\cite{Roy:2017mnz}. It must be noticed that our classification in thawing and freezing is based upon on the behavior of the solutions as described in the corresponding references, but such classifications cannot be read directly from the final form of $y_2$, as also a proper choice of initial conditions must be taken into account. More details can be found in Sec.~\ref{sec:gener-solut-equat} below.

\begin{table*}[htp!]
\caption{List of quintessence potentials and their corresponding
  closed form of $y_2$ in terms of the potential variables $y$ and
  $y_1$. In general, we see that $y_2/y$ is represented by a polynomial
  form in terms of the variable $y_1/y$. Also, it should be noticed
  that some of the potential free parameters, indicated here by the capital Latin letters, as in the case of the scale energy $A^4$, do not appear in the final form of $y_2$. The only dynamical parameters in the potentials, that end up in the final form of $y_2$, are indicated by the Greek letter $\lambda$ (although it should not be confused with the variable defined in Eq.~\eqref{eq:fundamental}). \label{tab:1}}
\begin{ruledtabular}
\begin{tabular}{|c|c|c|}
Ref. & Potential $V(\phi)$& Closed form of $y_2$ \\ 
\hline 
\hline 
\multicolumn{3}{|l|}{ Thawing potentials }  \\  
\hline 
\hline
\cite{linde1985universe,Roy:2014yta}& $A^4 (1 + B \phi)^{2 \lambda}$ & $\frac{1-\lambda}{2\lambda} \, y_1^2/y $ \\ 
\hline 
\cite{dutta2008hilltop} & $A^{4} \exp \left(- \phi^2/\lambda^2 \right)$ & $ \frac{12}{\kappa^2 \lambda^2} y - \frac{1}{2} y^2_1/y$ \\
\hline 
\cite{Freese:2014nla,Freese:1990rb}& $A^4 [1 + \cos(\phi/\lambda)]$ &
                                                                      $\frac{3}{\kappa^2 \lambda^2} y$  \\ 
                                                                      \hline                                                             \cite{peebles1988cosmology} & $A^{4+\lambda} \phi^{-\lambda}$ & $ -\frac{1}{\lambda} (\frac{\lambda}{2} + 1) y_1^2/y$ \\
                                        \hline                              \cite{Brax:1999yv} & $A^4 e^{2 \lambda \kappa^2 \phi^2}$ & $-24 \lambda y - \frac{1}{2} y_1^2/y$ \\

\hline 
\hline 
\multicolumn{3}{|l|}{Freezing potentials }  \\  
\hline 
\hline
\cite{starobinsky1980new,Carrasco:2015pla}& $A^4 (1 - e^{- \lambda \kappa \phi})^2$ & $- \sqrt{6} \lambda y_1$ \\ 
\hline 
\cite{Fang:2008fw} & $A^4 \cosh(\lambda \kappa \phi)$ & $ -6\lambda^2 y + \frac{1}{2} y_1^2/y$ \\ 
\hline 
\cite{Sahni:1999qe}& $A^4 [\cosh(\lambda \kappa \phi)]^{-1}$ &  $6 \lambda^2 y - \frac{3}{2} y_1^2/y$ \\ 
\hline 
\cite{UrenaLopez:2000aj} & $A^4 [\sinh(\lambda_1 \kappa \phi)]^{-\lambda_2}$ & $6  \lambda_1^2 \lambda_2 y  - (1/\lambda_2) (1 + \lambda_2/2) y_1^2/y$ \\ 
\hline 
\cite{Barreiro:1999zs}& $A^4 \left[ e^{\lambda_1 \kappa \phi} +
                        e^{\lambda_2 \kappa \phi} \right]$ & $6
                                                             \lambda_1
                                                             \lambda_2
                                                             y +
                                                             \sqrt{6}
                                                             (\lambda_1
                                                             +
                                                             \lambda_2)
                                                             y_1 +
                                                             \frac{1}{2}
                                                             y_1^2/y$ 

\end{tabular} 
\end{ruledtabular} 
\end{table*}

One can see that for the examples in Table~\ref{tab:1} the functional
forms of $y_2$ can be expressed in terms of the variables $y$ and
$y_1$, or more precisely, as a polynomial in terms of the ratio
$y_1/y$. It is then natural to consider that there exists a more
general function of $y_2$ in the form
\begin{equation} 
\label{GP}
y_2 = y \, \sum_{i = 0}^{n} \alpha_i \left( \frac{y_1}{y}
\right)^i \, .
\end{equation}
where $\alpha_i$ are constant coefficients. As also shown in the
examples in Table~\ref{tab:1}, the constant coefficients $\alpha_i$ will
then drive the dynamics of the quintessence field, although they will
not be directly related to other free parameters in the
potential, which are denoted with Latin capital letters in the examples of Table~\ref{tab:1}\footnote{We discuss in Appendix~\ref{sec:app-general} a more general approach for the functional form of the ratio $y_2/y$.}. 

For completeness, we show in Table~\ref{tab:2} the inverse process
that can be used upon Eq.~\eqref{GP} to obtain from it different quintessence potentials. The simplest possibility is $\alpha_j =0$, for which Eq.~\eqref{GP} can be written as $\partial^2_\phi V^{1/2} =0$ and then upon integration we obtain $V(\phi) = (A+B\phi)^2$, where $A,B$ are integration constants. This is precisely one particular example (with $\alpha_2 =0$) of Class Ia in Table~\ref{tab:2}.

In the most general case, Eq.~\eqref{GP} can be written as a differential equation by means of the definitions of the variables $y$, $y_1$ and $y_2$ in Eqs.~\eqref{eq:3}. Hence, 
\begin{equation}
\partial^2_{\kappa \phi} V^{1/2} + \frac{V^{1/2}}{12} \sum_{j=0} \alpha_j \left( -2\sqrt{6} \frac{\partial_{\kappa \phi} V^{1/2}}{V^{1/2}} \right)^j = 0 \, , \label{eq:diffeq}
\end{equation}
where the derivatives are calculated with respect to the dimensionless variable $\kappa \phi$. Using the auxiliary function $\lambda = y_1/y = -2\sqrt{6} \, \partial_{\kappa \phi} V^{1/2}/V^{1/2}$, we can write Eq.~\eqref{eq:diffeq} in the form
\begin{equation}
\partial_{\kappa\phi} \lambda = \frac{1}{2\sqrt{6}} \left[ \lambda^2 +  2 \sum_{j=0} \alpha_j \lambda^j \right] \, . \label{eq:fundamental}
\end{equation}
Thus, the inverse process to find the quintessence potential if the dynamical parameters $\alpha_j$ are given consists in the integration of the fundamental equation~\eqref{eq:fundamental}. In general terms, Eq.~\eqref{eq:fundamental} can be integrated by the method of partial fraction decomposition, for which we require first to find the roots of the polynomial on the right hand side. Once a solution is found for the auxiliary function $\lambda = \lambda(\kappa \phi)$, the corresponding quintessence potential is obtained from $V(\phi) = \exp\left[ -\lambda(\kappa \phi)/\sqrt{6} \right]$.

The cases in Table~\ref{tab:2} are those that correspond to the quadratic expansion ($\alpha_j =0$ for $j \geq 3$) in Eq.~\eqref{eq:fundamental}. It can be verified that there is a direct correspondence between the general cases in Table~\ref{tab:2} with the particular examples shown in Table~\ref{tab:1}, as long as the constants $A$, $B$ and $C$ are adjusted accordingly. From the numerical point of view, the most general form of the potential is obtained from $\alpha_j \neq 0$, and all other forms should be a subclass of this. But analytically this is not achievable as the integration scheme is different for different choices of $\alpha_j$, and this is why we find it more natural to classify the quintessence potentials in the four classes shown in Table~\ref{tab:2}.

\subsection{Dynamical and passive parameters \label{sec:dynamical-passive}}
We said before that the $\alpha$-parameters are the only dynamical ones, and then their allowed values suggest natural classifications of the potentials in general classes. We hereafter dub them dynamical parameters. Apart from this, there will be constants of integration ($A$, $B$ and $C$ in the examples of Table~\ref{tab:2}), which are then redundant from the dynamical point of view and do not have any influence of the behavior of the field solutions, except in the set up of the initial conditions. We will refer to them as passive parameters.

To briefly illustrate the difference with respect to the dynamical ones, let us consider the well known example in Class Ia which is the quadratic potential $V(\phi) = (m^2_\phi/2) \phi^2$ ($A=0$, $B=m_\phi/\sqrt{2}$ and $\alpha_2 = 0$), where $m_\phi$ is the mass of the scalar field\cite{Urena-Lopez:2015odd}. It can be shown that for this case $y_{1i} = 2\sqrt{2} B/H_i = 2m_\phi/H_i$, where $H_i$ is the initial value of the Hubble parameter (see also Sec.~\ref{sec:numer-analys-results} below). The initial values of the other two dynamical variables, $\theta_i$ and $\Omega_{\phi,i}$ must be fixed by taking additional considerations, like the expected contribution of the quintessence field at the present time. Thus, the value of parameter $B$ plays a role in the set up of the initial conditions of the field variables, even though it does not affect at all their evolution and dynamics. More about the free parameters in the potentials, both dynamical and passive, is discussed in the sections below.

\begin{table*}[htp!]
\caption{\label{tab:2} Examples of general quintessence potentials that are obtained from the reverse integration of the definition of the second potential variable $y_2$, see Eqs.~\eqref{GP} and~\eqref{eq:fundamental}. Notice that we only considered the expansion in Eq.~\eqref{GP} up to the second order, as the inverse process is not analytical if higher order terms are included.}
\begin{ruledtabular}
\begin{tabular}{|c|c|c|}
No & Structure of $y_2/y$ & Form of the potentials $V(\phi)$  \\ 
\hline 

Ia&  $\alpha_0 = 0, \alpha_1 = 0, \alpha_2 \neq  - \frac{1}{2}$ & $ (A  + B \phi)^{\frac{2}{(2 \alpha_2 +1)}}$ \\
\hline
Ib & $\alpha_0 = 0, \alpha_1 = 0, \alpha_2 = - \frac{1}{2}$ & $A^2 e^{2 B \phi}$ \\
\hline 

IIa & $\alpha_0 \neq 0, \alpha_1 = 0, \alpha_2 \neq - \frac{1}{2}$& $A^2 \cos \left[ \sqrt{\alpha_0 \kappa^2 (1 + 2 \alpha_2)} (\phi -  B) /2 \sqrt{3} \right]^{\frac{2}{1 + 2 \alpha_2}}$ \\
\hline 
IIb & $\alpha_0 \neq 0, \alpha_1 = 0, \alpha_2 = - \frac{1}{2}$ & $A^2 \exp \left({- \kappa^2 \alpha_0  \phi^2/12}) \exp({ 2 B \phi} \right)$ \\
\hline

IIIa & $\alpha_0 = 0, \alpha_1 \neq 0, \alpha_2 \neq - \frac{1}{2}$ & $\left[ A \exp \left( \alpha_1 \kappa \phi/\sqrt{6} \right) + B\right]^{\frac{2}{1+2 \alpha_2}}$ \\
\hline
IIIb & $\alpha_0 = 0, \alpha_1 \neq 0, \alpha_2 = - \frac{1}{2}$ & $A^2 \exp \left[ 2 B ~ \exp \left(	\kappa \alpha_1 \phi /\sqrt{6} \right) \right] $ \\
\hline

IVa & $\alpha_0 \neq 0, \alpha_1 \neq 0, \alpha_2 \neq - \frac{1}{2}$ & $A ^2 \exp(\frac{ \kappa \alpha_1 \phi}{ \sqrt{6} ( 1 + 2 \alpha_2)}) \left\{ \cos \left[ \left(- \frac{\kappa^2 \alpha_1 ^2}{24} + \frac{\kappa^2 \alpha_0}{12} (1 + 2 \alpha_2) \right)^{\frac{1}{2}} (\phi - B) \right] \right\}^{\frac{2}{1+2 \alpha_2 }}$ \\
\hline
IVb & $\alpha_0 \neq 0, \alpha_1 \neq 0, \alpha_2 = - \frac{1}{2}$ & $A^2 \exp \left[ \frac{\kappa \alpha_0 \phi}{\sqrt{6} \alpha_1	} + 2B \exp \left(	\frac{\kappa \alpha_1 \phi}{\sqrt{6}} \right) \right]$ 
\end{tabular}
\end{ruledtabular}
\end{table*}

\section{Approximate solution of the equations of motion \label{sec:gener-solut-equat}}
Here we will show how to obtain a solution of the equations of motion~\eqref{eq:10} that is of general applicability to any kind of quintessential potential of the  (monotonic) thawing type. This will in turn be useful also to obtain appropriate initial conditions for the general numerical solutions that will be used in Sec.~\ref{sec:numer-analys-results}.

It must be stressed out that the thawing condition for the quintessence models requires that, initially, the EoS is $w_{\phi} \simeq -1$ and also that $y_1 > 0$. This means that the quintessence EoS will deviate from the cosmological constant value at late times. Notice that the thawing condition is here chosen by hand, but our formalism also allows other possibilities (freezing, tracker, skater, etc.), which we leave for future studies.

We start by noting that from observations we expect the present value of the quintessence EoS to be $w_{\phi} \simeq -1$, which is equivalent in terms of the polar variables to $\theta < 1$. Moreover, at the epoch when the universe entered into the matter dominated phase from a radiation dominated phase, the scalar field energy density was still very subdominant $\Omega_{\phi} \ll 1$. By taking into account the approximations $\Omega_{\phi} \ll 1$ and $\theta \ll 1$, we neglect the second term in Eq.~\eqref{eq:10b} and find separate solutions during the radiation and matter domination eras. We shall then match the separate solutions at the time of the radiation-matter equality, and with this we shall try to make a reasonably good guess of the initial conditions of the universe which can lead to the present accelerated universe. 

One final note is that in the radiation and the matter dominated cases the $e$-foldings $N$ are different. For radiation domination $N_r = \ln(a/a_i)$ where $a_i$ is the initial value of the scale factor, whereas for matter domination $N_m = \ln(a/a_{eq})$, where $a_{eq}$ is the scale factor of the universe at the epoch of radiation-matter equality.

\subsection{Radiation dominated era \label{sec:radi-domin-era}}
As the universe is dominated by radiation the total EoS simply is
$\omega_{tot} = 1/3$, and due to the smallness of $\theta$ we can use
the following approximations: $\sin \theta \simeq \theta$ and $\cos
\theta \simeq 1$, so that also $w_\phi \simeq -1$. Hence,
Eqs.~\eqref{eq:10} reduce to 
\begin{eqnarray}
  \theta^{\prime} = - 3 \theta + y_1 \, , \quad y_1^{\prime} = 2 y_1
  \, , \quad \Omega_{\phi}^{\prime} = 4 \Omega_{\phi}\,
  , \label{eq:11}       
\end{eqnarray}
The growing solution of Eqs.~\eqref{eq:11}, within the radiation
domination era, are given by
\begin{equation}
  \theta_r = \theta_i (a/a_i)^2 \, , \; y_{1r} = y_{1i} (a/a_i)^2 \,
  , \; \Omega_{\phi r} =  \Omega_{\phi i } (a/a_i)^4 \,
  , \label{eq:12}
\end{equation}
where a subindex $r$ denote the solution during radiation domination
and a subindex $i$ denote the initial value of the corresponding
variable. In addition, we also find that $y_1 = 5 \theta$, which
is just the attractor solution for these variables during radiation domination.

\subsection{Matter dominated era \label{sec:matt-domin-era}}
As the universe is dominated by matter we now consider that
$\omega_{tot} = 0$, and after using the same approximations as in
Eqs.~\eqref{eq:11}, Eqs.~\eqref{eq:10} now become
\begin{equation}
  \theta^{\prime} = - 3 \theta + y_1 \, , \quad y_1^\prime =
  \frac{3}{2} y_1 \, , \quad \Omega_{\phi}^{\prime} = 3 \Omega_{\phi}
  \, . \label{eq:13}       
\end{equation}
After solving Eqs.~\eqref{eq:13} we obtain the matter dominated solutions
\begin{eqnarray}
  \theta_m &=& \left( \theta_{eq} - \frac{2}{9} y_{1eq} \right)
  (a/a_{eq})^{-3} + \frac{2}{9} y_{1eq} (a/a_{eq})^{3/2} \, ,
  \nonumber \\ 
  y_{1m} &=& y_{1eq} (a/a_{eq})^{3/2} \, , \; \Omega_{\phi m} =
  \Omega_{\phi eq} (a/a_{eq})^3 \, . \label{eq:14}     
\end{eqnarray}
Here, a subindex $m$ denote the solution during matter domination
and a subindex $eq$ denote the initial value of the corresponding
variable at the time of radiation-matter equality. In contrast to the previous radiation dominated case, we are not neglecting the decaying solution in Eq.~\eqref{eq:14} as it will be required below to handle the transition between the two cosmological eras.

We matched the approximate solutions~\eqref{eq:12} and~\eqref{eq:14}
at the time of radiation-matter equality $a_{eq} = \Omega_{r0}/\Omega_{m0}$ so that we can find a solution at matter domination that carries information about the initial conditions set up in radiation domination. From Eqs.~\eqref{eq:12} we find the values of the variables at radiation-matter equality: $\theta_{eq} = \theta_i (a_{eq}/a_i)^2$, $y_{1eq} = 5\theta_{eq} = y_{1i} (a_{eq}/a_i)^2$ and $\Omega_{\phi eq} = \Omega_{\phi i} (a_{eq}/a_i)^4$, which we substitute in Eqs.~\eqref{eq:14} to obtain
\begin{subequations}
\label{eq:matter-dom}
\begin{eqnarray}
\theta_m &=& \frac{10}{9} \left (\frac{a_{eq}}{a_i} \right) ^2\, \theta_i \left[ \left( \frac{a}{a_{eq}} \right)^{3/2} - \frac{1}{10}  \left( \frac{a}{a_{eq}} \right)^{-3} \right] \, , \label{eq:matter-doma} \\
y_{1m} &=& \frac{a^{1/2}_{eq}}{a^2_i} \, y_{1i} \, a^{3/2} \, , \label{eq:matter-domb} \\
\Omega_{\phi m} &=& \frac{a_{eq}}{a_i^4} \, \Omega_{\phi i}  \, a^3\, . \label{eq:matter-domc}
\end{eqnarray}
\end{subequations}

\subsection{Estimation of initial conditions \label{sec:estimation-of-init}}
Equations~\eqref{eq:matter-dom} can be used to estimate the initial conditions on the dynamical variables from the present values of  $\theta$ and $\Omega_\phi$. We will assume that the matter domination solutions~\eqref{eq:matter-dom} are valid up to the present time. This is not correct from a formal point of view, but we have verified the appropriateness of Eqs.~\eqref{eq:matter-dom} by a direct comparison with numerical solutions, see Sec.~\ref{sec:numer-analys-results} below. Hence, by taking $a=1$ in Eqs.~\eqref{eq:matter-dom}, the initial conditions for the quintessence dynamical variables can be estimated from
\begin{subequations}
\label{eq:7}
\begin{eqnarray}
  \theta_i &\simeq& \frac{9}{10} a^2_i \,
  \frac{\Omega^{1/2}_{m0}}{\Omega^{1/2}_{r0}} \theta_0
  \, , \label{eq:7a} \\
  \Omega_{\phi i} &\simeq& a_i^4 \, \frac{\Omega_{m0}}{\Omega_{r0}} \, \Omega_{\phi 0} \, , \label{eq:7b}
\end{eqnarray}
\end{subequations}
where we have only taken the leading term in Eq.~\eqref{eq:matter-doma} for the expression of $\theta_i$. The same procedure applied to Eq.~\eqref{eq:matter-domb} gives $y_{1i} = 5\theta_i$, which is exactly the attractor solution expected during radiation domination, see Eqs.~\eqref{eq:12}.

\subsection{Final considerations \label{sec:final-cons-}}
We learn from the first of Eqs.~\eqref{eq:7}, because of the direct relation
of the polar variable $\theta$ to the scalar field EoS, that the
present value of the latter $w_{\phi 0}$ is an output value that is
directly determined by the initial value $\theta_i$. This would be in
agreement with the standard field approach to quintessence, in which
one has to try different initial conditions, for both $\phi_i$ and
$\dot{\phi}_i$, to explore a given range of values for
$w_{\phi0}$. The main difficulty in the latter is that there is
not a straightforward relation between the pair $(\phi_i,\dot{\phi}_i)$ and the present value $w_{\phi0}$, and then the search of initial
conditions for a proper sampling of $w_{\phi0}$ must be done
differently for each potential $V(\phi)$. One big advantage of our
approach in this respect is that we can avoid such a hassle and use generic initial conditions for all cases, irrespectively of the particular form of the potential.

On the other hand, the relation between the present and initial values
of the scalar field density parameter $\Omega_\phi$ is just the one
that is obtained for a cosmological constant; this is hardly
surprising as one assumption was that the scalar field EoS was close
to $-1$ for most of the evolution of the Universe.

One final quantity of interest is the ratio $y_1/y$ at the present time; from the solutions presented above we find that 
\begin{equation}
  \frac{y_{10}}{y_0} \simeq \frac{y_{10}}{\Omega^{1/2}_{\phi 0}} =
  \frac{y_{1i}}{\Omega^{1/2}_{\phi i}} \simeq \frac{9}{2} \left[ \frac{2(1
      + w_{\phi 0})}{\Omega_{\phi 0}} \right]^{1/2} \, . \label{eq:8}
\end{equation}
Its present value is basically set up by the initial conditions, or equivalently, as suggested by the last equality in Eq.~\eqref{eq:8}, by the present values of the quintessence parameters. Hence, the ratio $y_1/y$ should remain small for most of the evolution of the Universe if the quintessence field is to be the dark energy, i.e. if $1+w_{\phi 0} \sim 0$. This reinforces our assumption that it is just enough to consider an expansion up to the second order in Eq.~\eqref{GP}, and then the dynamics of the quintessence fields will be represented, in general, by the values of the first three coefficients $\alpha_0$, $\alpha_1$ and $\alpha_2$.

\section{General properties of quintessence models \label{sec:numer-analys-results}}
In this section, we shall study the evolution of the universe
considering the general form of $y_2 = \alpha_0 y + \alpha_1 y_{1} +
\alpha_2 y_{1}^2/y$, and we shall explain the general
procedure to constrain the dynamical parameters $\alpha$.

\subsection{Class Ia and the cosmological constant \label{sec:classIa-cosmo}}
We explain here the correspondence, within our approach, between quintessence and the cosmological constant. Let us start with the simplest possibility which is $y_2 =0$, in terms of the expansion of $y_2$ in Eq.~\eqref{GP} this also means that:
$\alpha_0 = \alpha_1 = \alpha_2 =0$. As shown in Table~\ref{tab:2} this case corresponds to the parabolic potential $V(\phi) = (A +
B\phi)^2$, where $A$ and $B$ are integration constants. The potential has its minimum at $\phi_c = -A/B$, and then the mass of the quintessence field is simply given by $m^2_\phi \equiv \partial^2_\phi V (\phi_c) = 2B^2$. Thus, parameter $B$ gives us information about the mass scale of the quintessence field, whereas parameter $A$ tells us about the displacement of the minimum away from the origin at $\phi=0$. Moreover, there is now a straightforward interpretation of one of the potential parameters: $y_1 = 2 \sqrt{2} B/H$, see Eq.~\eqref{eq:3b}, and then we find that at all times $y_1 = 2m_\phi/H$. The parameter $A$ is left undetermined as it plays no role in the dynamics of the quintessence field.

  Let us in addition impose $B=0$, which also means that $y_1 =0$ for the whole evolution. Equation~\eqref{eq:4b} is identically satisfied, whereas Eq.~\eqref{eq:4a} provides the solution $\tan(\theta/2) = \tan(\theta_i/2) \, (a/a_i)^{-3}$; we find that $\theta \to 0$ as $(a/a_i) \to \infty$, and then also that $w_\phi \to -1$ at late times. This case corresponds to the case $V(\phi) = \mathrm{const.}$, that is, to the so-called skater models discussed in\cite{Linder:2005nh,Sahlen:2006dn}. Skater models then belong to our Class Ia of quintessence potentials under the condition $y_1 =0$.\footnote{\bf The evolution equation for the variable $\theta$ in this case simply is $\theta^\prime = -3 \sin \theta$, which in terms of the EoS can be rewritten as $w^\prime_\phi = -3 (1-w^2_\phi)$. The foregoing equation is exactly the one for skater models\cite{Linder:2005nh}.}

 We now revise the case of a constant EoS $w_{\phi} = w_{\phi 0}$. Although this can be seen as the simplest generalization of the cosmological constant, from Eqs.~\eqref{eq:10} we find that a constant EoS in the quintessence case could be obtained if, apart from $y_1 = 0$, we also impose $\theta_i = - \cos^{-1} (-w_{\phi 0})$ and $\theta^\prime = 0$. However, the latter condition cannot be sustained if $\theta \neq 0$, see Eq.~\eqref{eq:10a}, and then we get the same situation for skater models described in the above paragraph in which $\theta \to 0$. That is, the quintessence EoS cannot remain constant and must evolve towards the value $w_\phi \to -1$. These same calculations show that the only consistent conditions for a constant EoS are $\theta =0$ and $\theta^\prime =0$, which correspond exactly to the cosmological constant case.

 The morale from this discussion is then twofold. First, it is not possible to find a quintessence solution that emulates a constant EoS for DE apart from the cosmological constant case. And second, in terms of the parameters in our approach, the cosmological constant is just the null hypothesis ($\theta =0$ and $y_1 =0$), and then any deviation from the null value of the dynamical variables and parameters will be a measurement of the preference of the data catalogs on the quintessence models.

\subsection{The quintessence EoS \label{sec:quintessence-equation}}
One of the primary cosmological parameters in the studies of DE is the EoS of the DE field. In our approach, the EoS is, through the relation $w_\phi = -\cos \theta$, also one of the dynamical variables to describe the evolution of the quintessence field. Here we will discuss the influence of the dynamical parameters $\alpha$ and in doing so we will determine the behavior of the EoS under general quintessence potentials.

\begin{figure}[ht!] 
\centering
\includegraphics[width=0.49\textwidth]{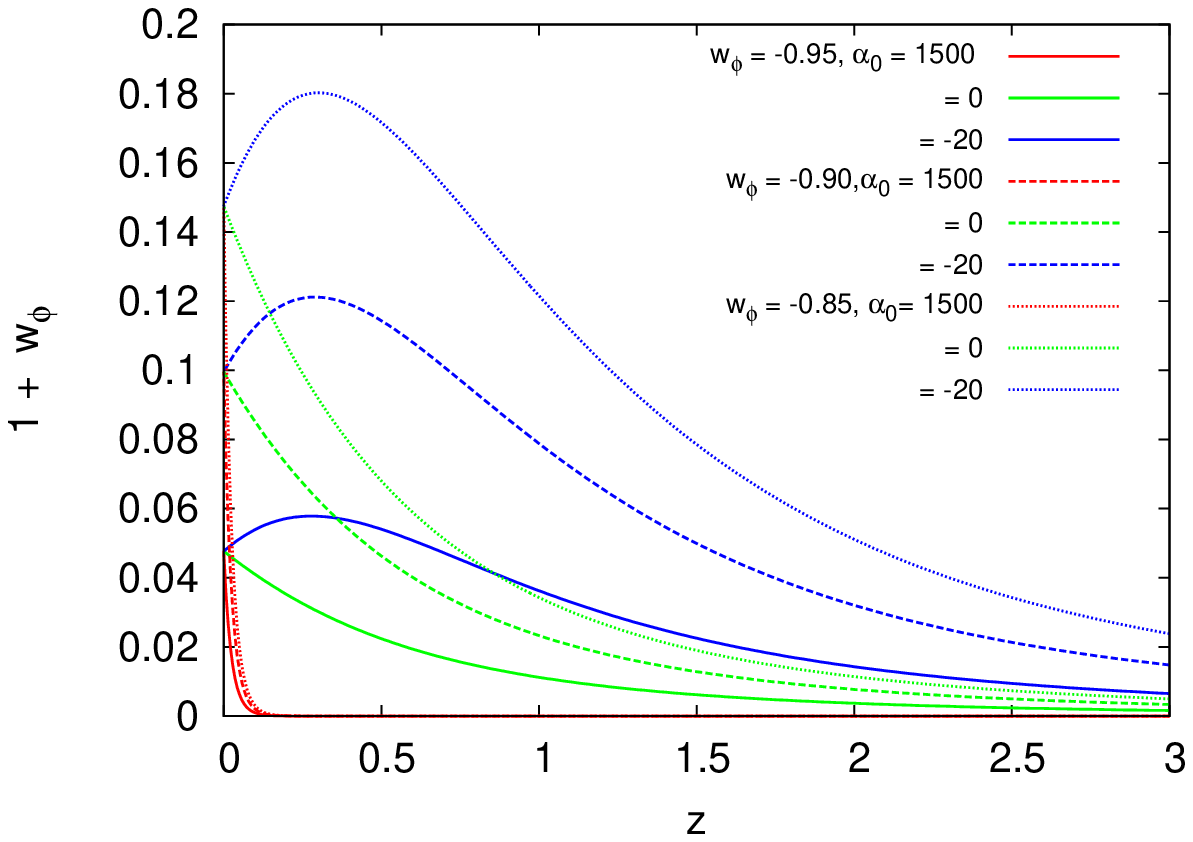} 
\includegraphics[width=0.49\textwidth]{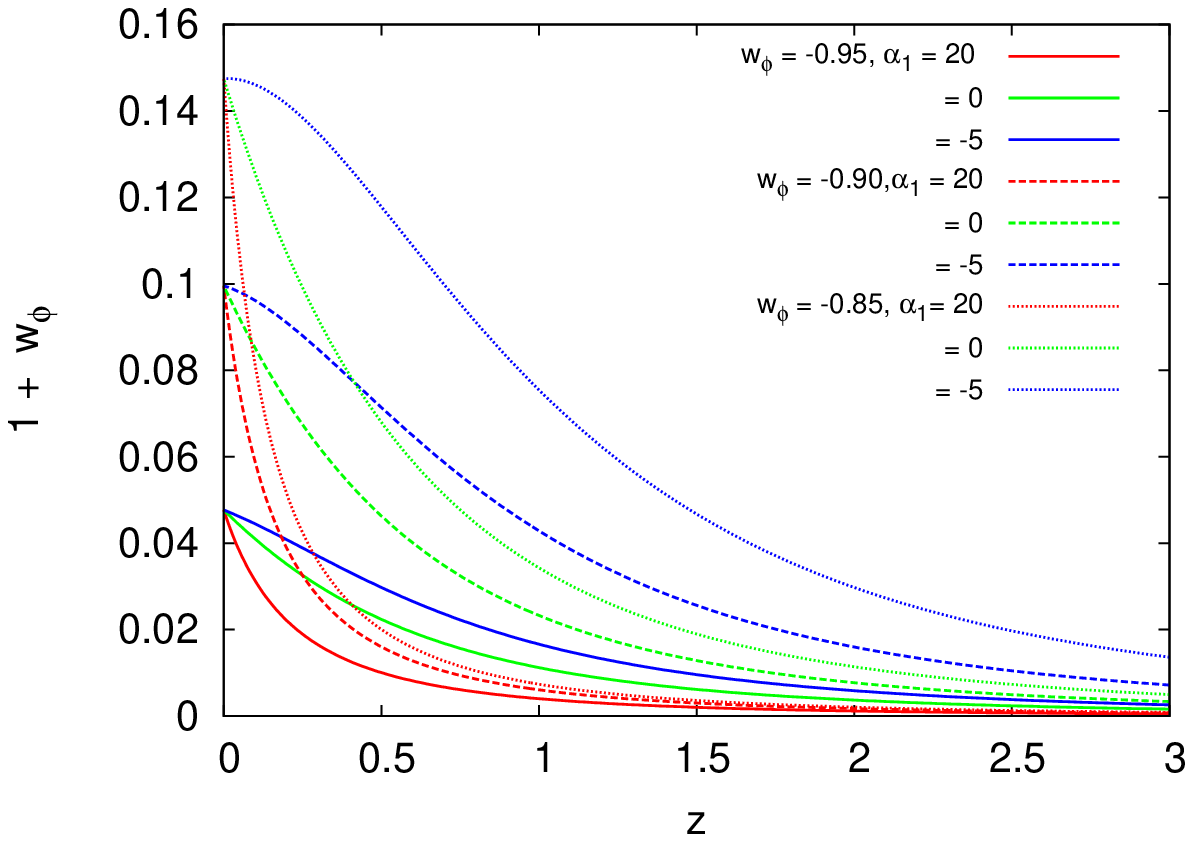}
\includegraphics[width=0.49\textwidth]{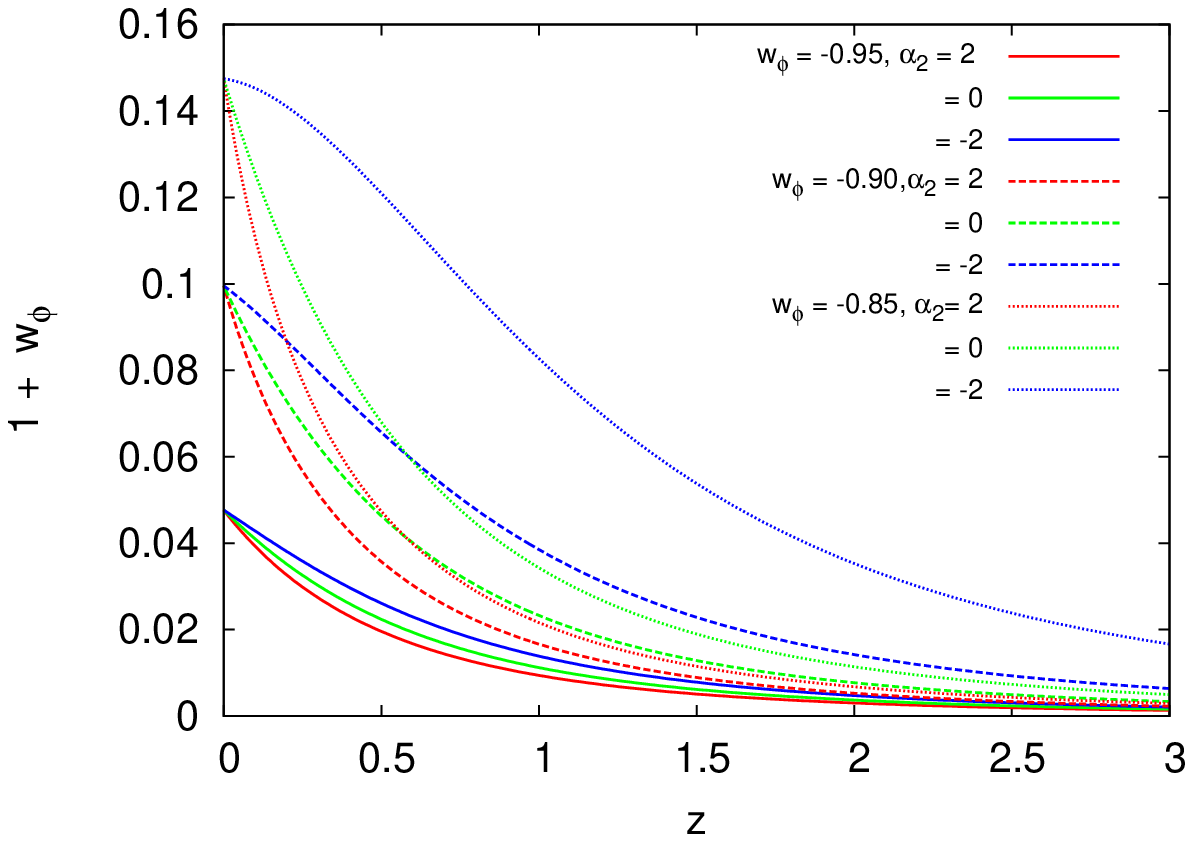}
\caption{\label{fig:wphi} Plots of $1 + w_{\phi} $ as a function of the redshift $z$ for the values indicated of the dynamical parameters $\alpha_0$ (top), $\alpha_1$ (middle) and $\alpha_2$ (bottom). Notice that the curves deviate from the cosmological constant value ($w=-1$) for $z < 3$, in general the curves grow monotonically as $z \to 0$ but a small bump appears if the dynamical parameters take on negative values.}
\end{figure}

We show in Fig.~\ref{fig:wphi} the plots of $1 + w_{\phi}$ as a function of the redshift $z$ for different values of the dynamical parameters $\alpha_0, \alpha_1$ and $\alpha_2$, respectively.  The plots are for the quantity $1 + \omega_{\phi} $ instead of $\omega_\phi$ as we are interested about the deviations of the quintessence EoS from the cosmological constant case. But also because at present we can make the approximation $1 + w_{\phi 0} \simeq \theta^2_0/2$, and we see that it is variable $\theta$ that provides such deviations. The numerical solutions are grouped according to the Class in Table~\ref{tab:2} they belong to and for the indicated values of the dynamical parameters. 

From these plots, one can clearly see that the variation of the EoS is more sensitive to $\alpha_2$, and less sensitive to the variation of $\alpha_0$. This is just the expected result as $\alpha_2$ is the partner coefficient of $y^2_1/y^2$ in the series expansion~\eqref{eq:12}.  It is interesting to note from Fig.~\ref{fig:wphi} that a desired value of EoS of the dark energy can be obtained for a wide range of $\alpha$ parameters. Recent cosmological observations can only constrain the present value of the EOS, but unless there is any constraint on the evolution of the EOS it will not be possible to choose from the solutions for different $\alpha$ parameters. Hence, our expectation is that the statistical analysis using cosmological observations in the next section will not be able to constrain the $\alpha$ parameters. Additionally, we also see that the curves show a monotonic growth if the dynamical parameters are positive, but the curves develop a bump (ie a maximum appears) if the parameters are negative enough. We can only speculate that this latter effect seems to be an indication for the possible appearance of oscillations in the evolution of the EoS, but we will leave this topic for a future study.

As for the monotonic growing behavior at late times that is found for positive values of the dynamical parameters, it can be fit by the following expression,

\begin{equation}
1 + w_\phi = \left( 1+w_{\phi 0} \right) a^3 \left[ w_1 + (1- w_1) a^\gamma \right] \, , \label{eq:EoS-param}
\end{equation}
where $w_1$  and $\gamma$ are free parameters. Notice that $w_{\phi 0}$  is explicitly present in Eq.~\eqref{eq:EoS-param} to ensure that the actual value of the EoS is obtained at $a=1$. The second term on the rhs of Eq.~\eqref{eq:EoS-param} corresponds to the expected behavior during matter domination: from the leading solution in Eq.~\eqref{eq:matter-doma}, and together with Eq.~\eqref{eq:7}, we obtain that $(1+w_\phi)_m \simeq (1/2) \theta^2_m \simeq (1/2) \theta^2_0 a^3$. Hence, for those cases in which this approximation is good enough until the present time we expect that $w_1 \sim 1$ and $\gamma \sim 0$. Any difference with respect to these values will signal the transition from matter to quintessence domination and of the presence of the dynamical parameters $\alpha$.
 
\begin{figure*}[ht!] 
\centering
\includegraphics[width=0.49\textwidth]{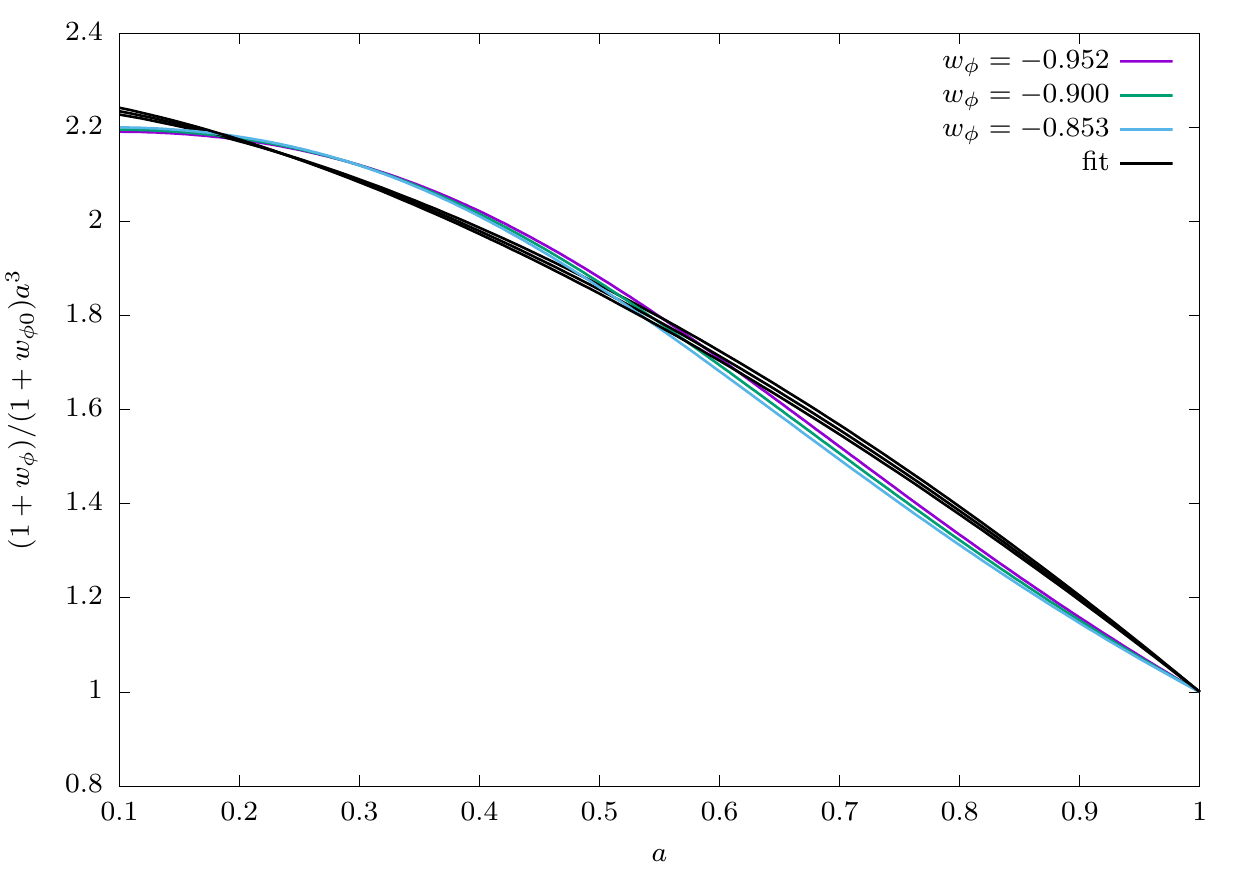}
\includegraphics[width=0.49\textwidth]{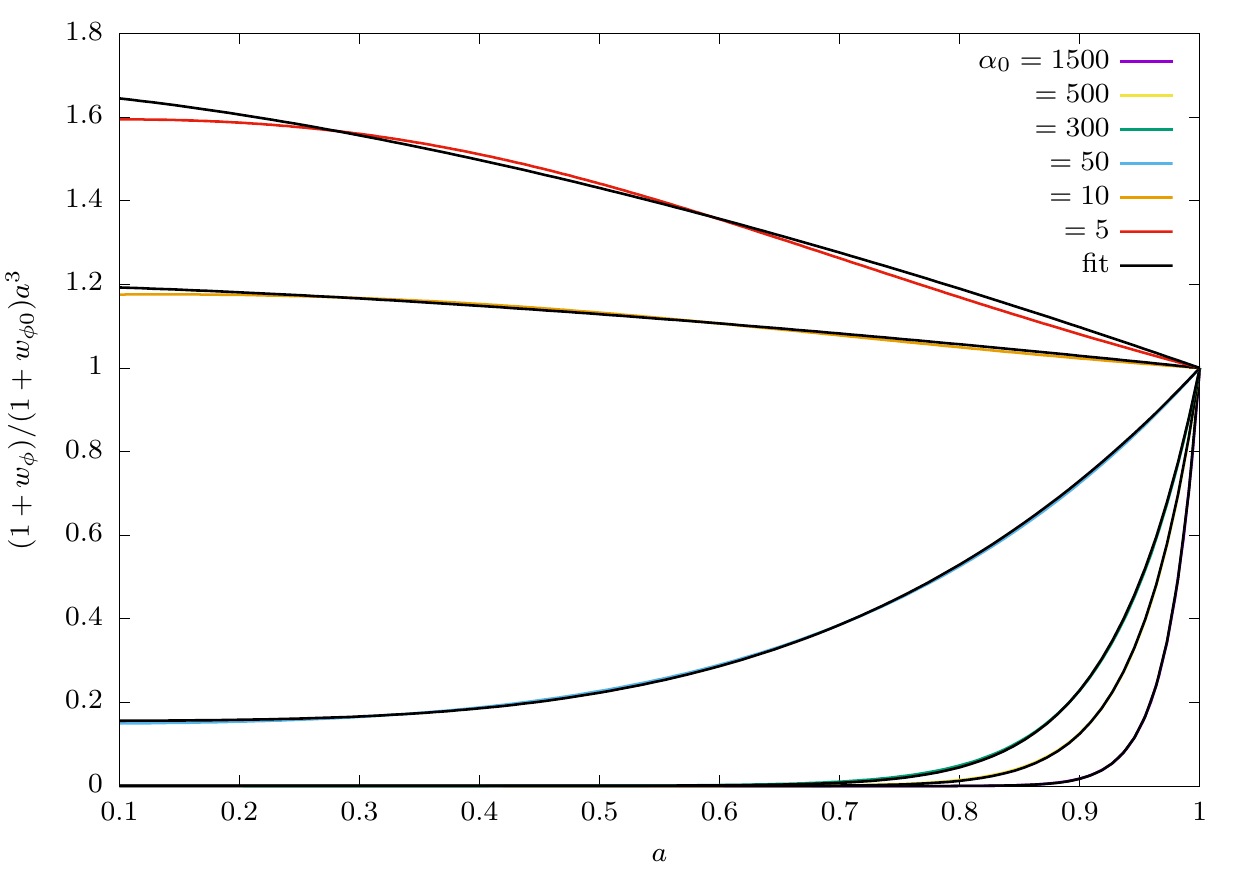}
\includegraphics[width=0.49\textwidth]{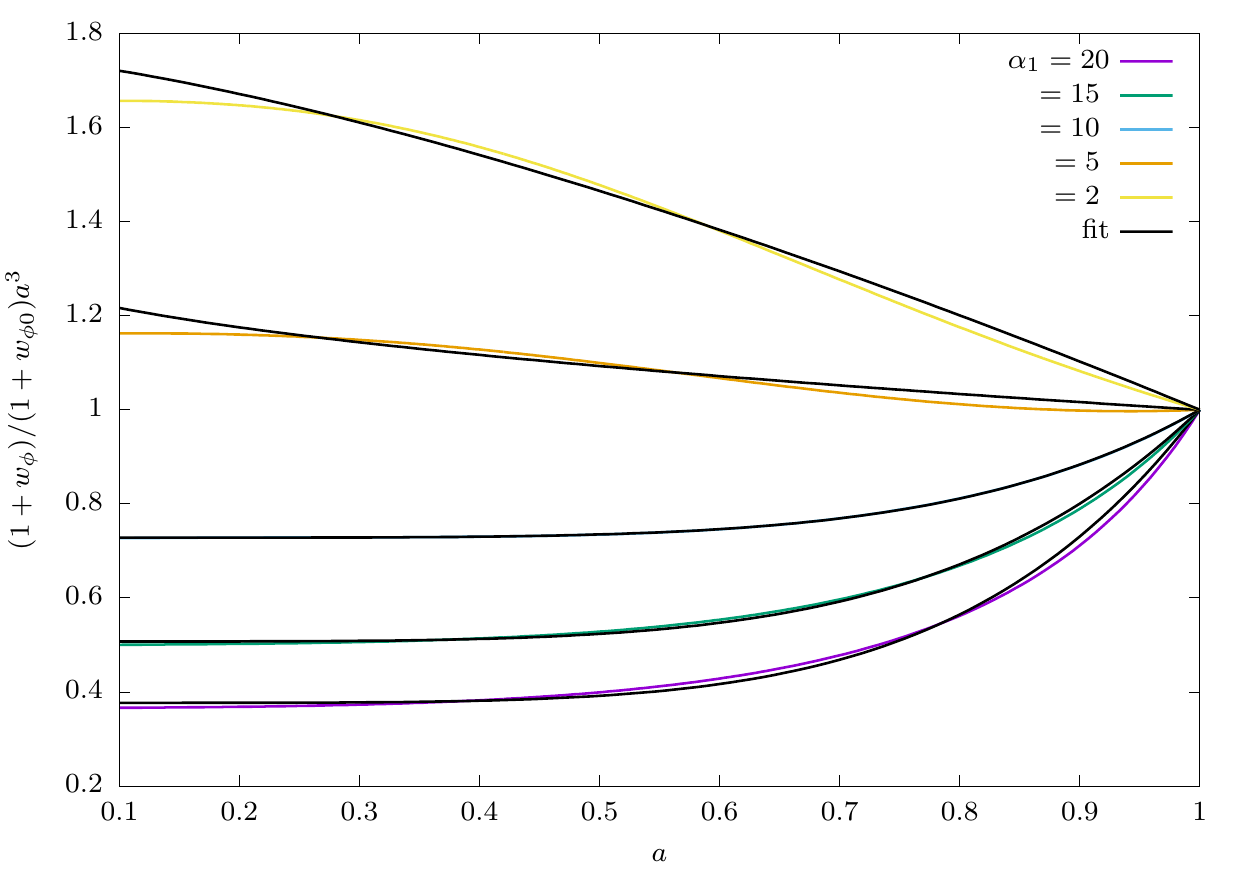}
\includegraphics[width=0.49\textwidth]{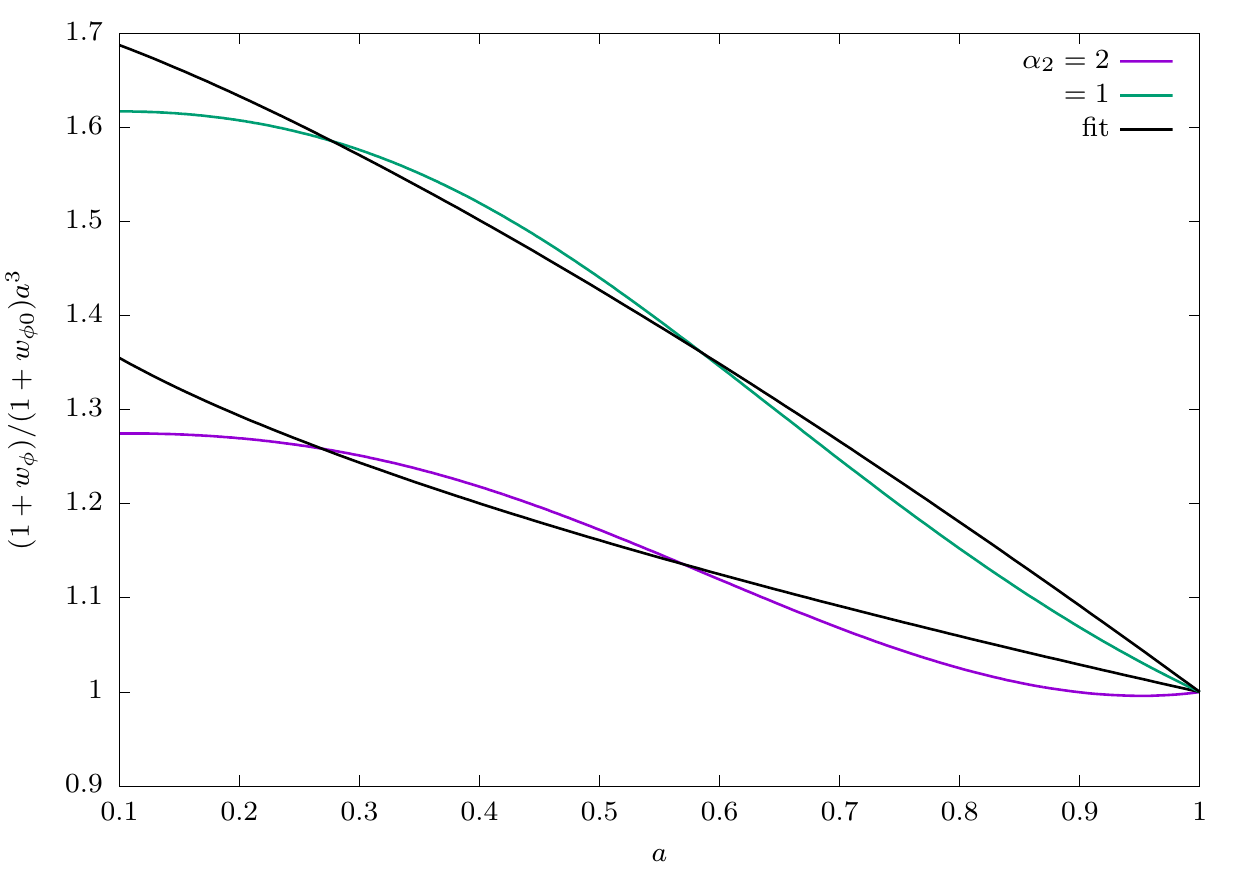}
\caption{\label{fig:fit} Fitting of the proposed parametrization in equation (\ref{eq:EoS-param}) to the numerical solutions obtained using the CLASS code corresponding to the Table \ref{tab:6}. These plots are for ${(1 + w_{\phi})}/(1 + w_{\phi 0}) a^3 $ as a function of the scale factor $a$ in the range [0.1 : 1]. Top-left plot corresponds to the Class Ia where $\alpha_i = 0$. Top-right corresponds to Class II where $\alpha_1 = \alpha_2 = 0$. In bottom-left the the plots are for the Class IIIa, $\alpha_0 = \alpha_2 = 0$ and in bottom-right the plots are for $\alpha_0 = \alpha_1 = 0$ which belongs to Class I.}
\end{figure*}

The results from a least-squares fitting of the parametrization~\eqref{eq:EoS-param} to the numerical solutions obtained from CLASS in some selected cases are shown in Table~\ref{tab:6} and in Fig.~\ref{fig:fit}, in all examples we considered the scale factor $a$ in the range $[0.1:1]$. It can be verified that the fits are indeed very good in all cases as the standard errors around the obtained values of the parameters are $\lesssim 1\%$. Not surprisingly, it is consistently found that $\gamma \gtrsim 0$, which indicates that the EoS accelerates its growth from $-1$ as the quintessence field starts to dominate the matter budget. 

\begin{table}[tp!]
\caption{\label{tab:6} The values of the parameters $\gamma$ and $w_1$ obtained from a least-squares fit of the parametrization~\eqref{eq:EoS-param} to some of the numerical solutions in Fig.~\ref{fig:wphi}. Notice that in general $\gamma \gtrsim 1$, which means that the leading power in the parametrization~\eqref{eq:EoS-param} is larger than $a^3$. The standard errors around the obtained values of the parameters are $\lesssim 1\%$.}
\begin{ruledtabular} 
\centering
\begin{tabular}{|c|c|c|}
\multicolumn{3}{|l|}{Class Ia: $\alpha_0 = \alpha_1 = \alpha_2 =0$}  \\ 
\hline 
$\omega_{\phi 0}$ & $\gamma$  & $w_1$ \\
 \hline
$-0.952$  & $ 1.691 \pm 0.016$ & $2.253 \pm 0.003$ \\
\hline 
$-0.900$  & $1.627 \pm 0.016$ & $2.264 \pm 0.004$ \\
\hline  
$-0.853$ & $1.570 \pm 0.017 $ & $2.276 \pm 0.004 $ \\
\hline 
\multicolumn{3}{|l|}{Class II: $\alpha_1 = \alpha_2 =0$, and $w_{\phi 0} = -0.853$}  \\ 
\hline 
 $\alpha_0$ & $\gamma$ & $w_1$ \\
\hline
1500  &  $38.481 \pm 0.005$  & $3.219 \times 10^{-5} \pm  9.566 \times 10^{-6} $ \\
\hline
500 & $ 19.738 \pm 0.013 $ & $ 3.367 \times 10^{-4} \pm 6.042 \times 10^{-5}$ \\
\hline 
300 & $13.998 \pm 0.017$ & $ 1.065 \times 10^{-3} \pm 1.317 \times 10^{-4}$  \\
\hline
50 & $ 3.648 \pm 0.006$ & $ 0.156  \pm 3.354 \times 10^{-4}$ \\
\hline
10 & $ 1.500 \pm 0.020 $ & $1.198 \pm 0.001 $ \\
\hline 
5 & $1.500 \pm 0.017$ & $ 1.665 \pm 0.003$ \\
\hline 
\multicolumn{3}{|l|}{Class IIIa: $\alpha_0  = \alpha_3 =0$, and $w_{\phi 0} = -0.853$}  \\ 
\hline
\hline 
 $\alpha_1$ &$\gamma$ & $w_1$ \\
 \hline
20 & $5.373 \pm 0.034 $ & $0.379 \pm 0.0007$ \\
\hline 
15 & $ 4.928 \pm 0.022$ & $0.507 \pm 0.0004$ \\
\hline 
10 & $5.312 \pm 0.002$ & $0.728 \pm 1.27 \times 10^{-5}$
\\
\hline
5 & $0.503 \pm 0.048$ & $1.315 \pm 0.0223$ \\
\hline 
2 & $1.393 \pm 0.018$ & $ 1.751 \pm 0.019$ \\
\hline 
\multicolumn{3}{|l|}{Class I: $\alpha_0 = \alpha_2 =0$, and $w_{\phi 0}  = -0.853$} \\
\hline
 $\alpha_2$ &$\gamma$ & $w_1$ \\
 \hline
2 & $ 0.599 \pm 0.046 $ & $1.474 \pm 0.025$ \\
\hline
1 & $1.282 \pm 0.021$ & $1.725 \pm 0.006$
\end{tabular}
\end{ruledtabular}
\end{table}

Equation~\eqref{eq:EoS-param} can be compared with other parameterizations of the dark energy EoS, like the famous Chevalier-Polarski-Linder one: $w = w_0 + w_1(1-a)$, which is clearly inappropriate to describe the evolution of the quintessence models in this work. There exist other parameterizations, see for instance\cite{Jaber:2017bpx,delaMacorra:2015aqf,Akarsu:2015yea,Escamilla-Rivera:2016aca,Zhao:2017cud,Jaime:2018ftn} and references therein, but they usually have a more complicated form than Eq.~\eqref{eq:EoS-param}. Although they may serve to test more complicated DE models, they are certainly not the best options to test one DE model as simple as quintessence.

Like in the case of the CPL parametrization, notice that Eq.~\eqref{eq:EoS-param} uses the present value of the EoS as an explicit parameter, but one clear advantage of our approach is that we are parameterizing the underlying dynamical variable $\theta$, and then we are recovering the right behavior of the EoS at early times. Notwithstanding this, we will not pursue a study of the dynamics represented by the parametrized EoS~\eqref{eq:EoS-param} because of the obvious degeneracies with the dynamical parameters: one can see from Table~\ref{tab:6} that different combinations of the $\alpha$'s will result in similar values of the free parameters $\gamma$ and $w_1$. Also, our parameterization~\eqref{eq:EoS-param} is only valid for redshifts $z \lesssim 10$, as for larger redshifts we need to take into account the full solutions for radiation domination and the radiation-matter transition, see Sec.~\ref{sec:gener-solut-equat} above. All of this makes any reconstruction of the quintessence potential from the EoS parametrization fruitless, and then it is more convenient to work directly with the dynamical variables $\alpha$ extracted from the potentials.

\section{Observational constraints and results \label{sec:Observa-const}}
Here we discuss our general strategy to put observational constraints on the dynamical parameters that characterize the quintessence field. 

\subsection{General setup and datasets}
We use an amended version of the Boltzmann code CLASS \cite{lesgourgues2011cosmic} and the Monte Carlo code Monte Python \cite{Brinckmann:2018cvx,Audren:2012wb}. Amendments to CLASS includes those necessary for MontePhyton to be able to sample the parameters that we describe next. 
There are 6 parameters that we want to constrain: $\theta_0, y_{10}, \Omega_{\phi 0}, \alpha_0, \alpha_1, \alpha_2$, but only 5 of them are required as input parameters, namely: $\Omega_{\phi 0}, \theta_0, \alpha_0, \alpha_1, \alpha_2$, because the value of $y_{10}$ is to be inferred from the full numerical evolution. It must be stressed out that as we sample the values of $\theta_0$ we will also be sampling the present values of the quintessence EoS $w_{\phi 0}$ through the relation $w_{\phi 0} = - \cos \theta_0$. In practice, the present EoS is then an input value and we will have full control of its sampling, which is another advantage of our method and variables over the standard approach to quintessence fields.

As in many other instances, we still need to finely tune the initial
values of the dynamical variables at the beginning of every numerical
run. For that, we write $y_{1i} = 5\theta_i$, $\theta_i = P
\times$Eq.~\eqref{eq:7a} and $\Omega_{\phi i} = Q
\times$Eq.~\eqref{eq:7b}, where the values of $P$ and $Q$ are
adjusted with the shooting method already implemented within CLASS for
scalar field models. A few iterations of the shooting routine are enough to find the correct values of $\theta_i$, $y_{1i}$ and
$\Omega_{\phi i}$ that lead to the desired $\Omega_{\phi 0}$ and
$w_{\phi 0}$ with a very high precision; in all instances it has been found 
that $P,Q = \mathcal{O}(1)$, which indicates that Eqs.~\eqref{eq:7}
are good approximations to the required initial conditions. Here we only consider the background dynamics of the quintessence fields and leave the study of their linear perturbations for a future work.

In doing a full sampling of the dynamical parameters $\alpha_0, \alpha_1, \alpha_2$, we will also be sampling the general form of the potentials shown in Table~\ref{tab:2}. This way we expect to be able to impose constrains on the dynamical parameters but not on the passive ones of the potential $V(\phi)$. As explained before, these other parameters are related and can obtained from the dynamical variables $\theta_0$, $y_{10}$ and $\Omega_{\phi0}$, although this would have to be done case by case for each one of the potentials in Table~\ref{tab:2}. For purposes of generality, we will focus on the constraints to the dynamical parameters and consider only two examples of constraints on passive parameters.

We use two data sets that are sensitive to the background quantities: (i) the SDSS-II/SNLS3 Joint Light-curve Analysis (JLA) supernova data\cite{Betoule:2014frx}  and (ii) BAO measurements (Barionic acoustic oscillations), in this case the following datasets are included in the likelihood: 2dFGS,MGS, DR11 LOWZ and DR11 CMASS \cite{Anderson:2013zyy}. We imposed a Planck2015 prior on the the baryonic and cold dark matter contribution\cite{Adam:2015rua, Aghanim:2015xee, Ade:2015rim,Ade:2015xua,Ade:2015lrj}: $\omega_b = 0.02230 \pm 0.00014 $ and $\omega_{cdm} = 0.1188 \pm 0.0010$; whereas for the scalar field parameter we used flat priors in the range $-20<\alpha_0<20$,  $-5<\alpha_1<5$ and $-2<\alpha_2<2$. 
The total set of parameters being sampled are: $\omega_b$, $\omega_b$, $H_0$, and the scalar field contribution $\Omega_{\phi0}$ is set by the closure relation for the given $\theta_0$, $\alpha_0$, $\alpha_1$ and $\alpha_2$;  whereas the set of derived parameters is: $\Omega_m$, $\Omega_{\phi}$, $w_{\phi}$ and $y_1$.

\subsection{General results \label{sec:general-results}}
 
\begin{figure*}[h!] 
\centering
\includegraphics[width=\textwidth]{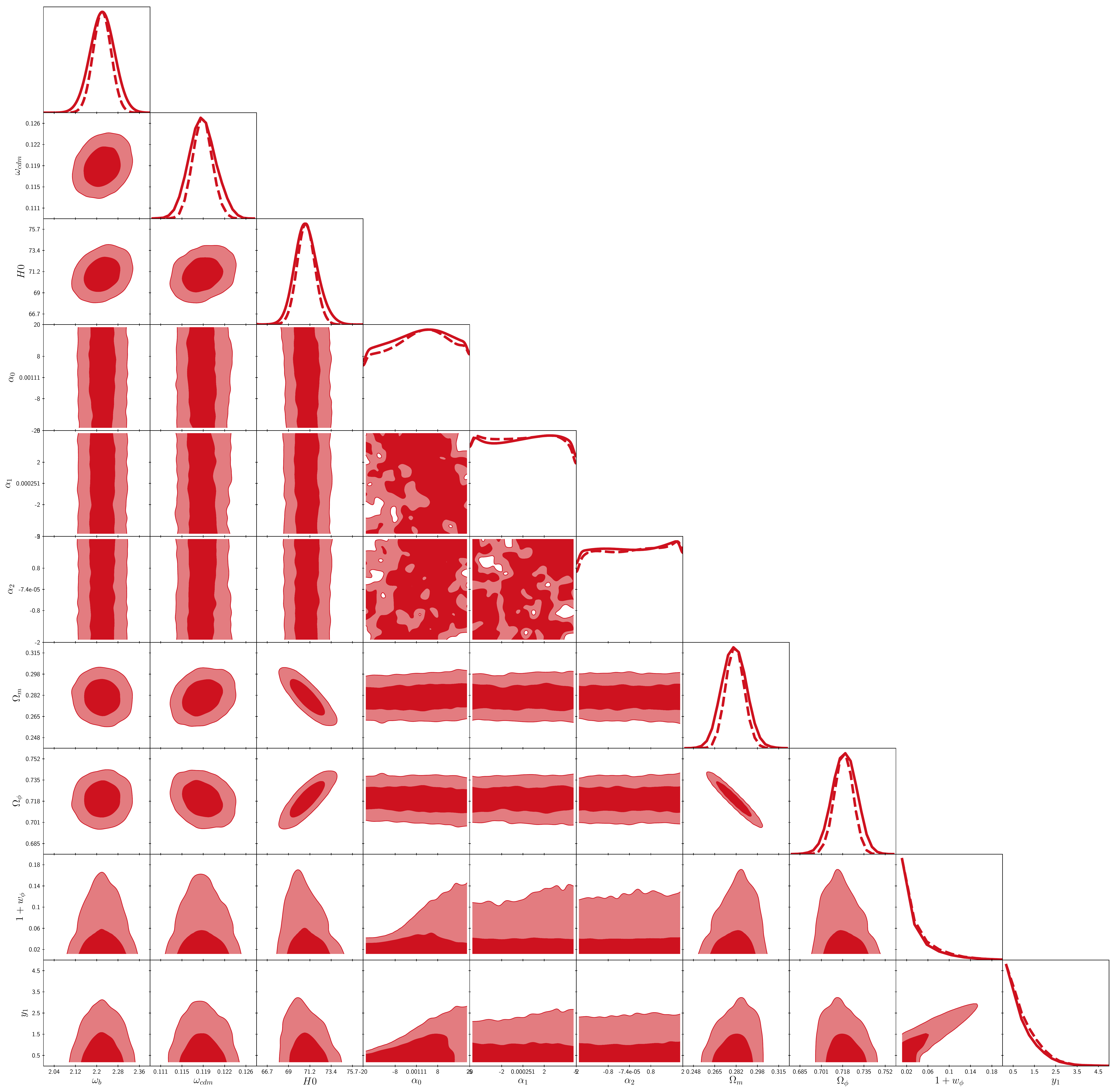}
\caption{\label{fig:triangle} Posterior distributions (solid lines) and mean likelihood (dashed line) for the constrained cosmological parameters using BAO+JLA dataset plus a PLANCK15 prior.
The data sets considered does not show any preference for a particular quintessence potential. See the text, Sec.~\ref{sec:numer-analys-results}, for more details. }
\end{figure*}

 The general constraints on the parameters of the quintessence models are shown in Fig.~\ref{fig:triangle}, where the 1D and 2D posterior distributions are represented in a triangle plot;it is also shown the Mean Likelihood Estimate (MELE) (dashed lines), which is another output from the Montepython code. In what follows we report our results using the median values of the 1D Posterior (not the mean likelihood) plus/minus a confidence interval, which is defined as the range containing 90\% of the samples. This particular choice is because some of the posterior parameters are not Gaussian. All of our analyses achieved a convergence ratio (Gelman-Rubin criteria) of $R-1 \approx 0.005$ for the standard cosmological parameters, although for some of the scalar field parameters the best convergence ratio we got did not go below $R-1 \approx 0.01$.  
Notice that all parameters are well constrained to some region of the parameter space, in both the posteriors and the MELE, except for the dynamical parameters $\alpha$ whose posterior distributions are plainly flat along the full prior range. This means that the data sets considered does not show any preference for a particular quintessence potential.

The present contribution of the quintessence field to the matter budget results in $\Omega_{\phi 0} = 0.719^{+ 0.015} _{-0.015}$, which is in agreement with previous studies\cite{ade2016planck}. 
Similarly, we find that  the EoS $1 + w_{\phi 0} < 0.107$ and that $0 <  y_1 < 2.24$  (95\% C.L.), values that are in agreement with the cosmological constant value $w_\phi =-1$ and $y_1 =0$. These results together show that the quintessence models revolve around the cosmological constant values. 

We now go back to the flat posteriors of the dynamical parameters $\alpha$. It means that a solution of the quintessence field compatible with the observational dataset can always be found for any value of the dynamical parameters, and the reason behind such result is that the initial conditions of the quintessence variables can be finely tuned accordingly to compensate for any $\alpha \neq 0$. For instance, for larger values of any of the dynamical parameters we can start the field evolution closer to the cosmological constant case, so that initially $w_\phi \to -1$ as much as necessary.

Moreover, the flat posteriors in Fig.~\ref{fig:triangle} also imply that there is not clear preference for any of the classes of potentials in Table~\ref{tab:2}. Given this situation, it may be reasonable to just consider the most economic possibility which is Class Ia in Table~\ref{tab:2}: $V(\phi) = (A + B\phi)^2$. As discussed in Sec.~\ref{sec:dynamical-passive}, one actually recover the quadratic potential if $A=0$ and $B = m_\phi/ \sqrt{2}$, and then we can say that for practical purposes no quintessence potential can fit the data any better than the quadratic potential.

We now turn our attention to the passive parameters in the quintessence potentials. In contrast to dynamical parameters, we shall argue that the passive ones can be subjected to observational constraints. It must be noticed that passive parameters, in their role as integration constants in Table~\ref{tab:2}, can only be determined if we fix either the initial or the final conditions in the solutions of the equations of motion. In the cosmological context we are interested in the final conditions as it is necessary to adjust the parameters to recover the present values of different observables.

\begin{figure}[th!] 
\centering
\includegraphics[width=0.49\textwidth]{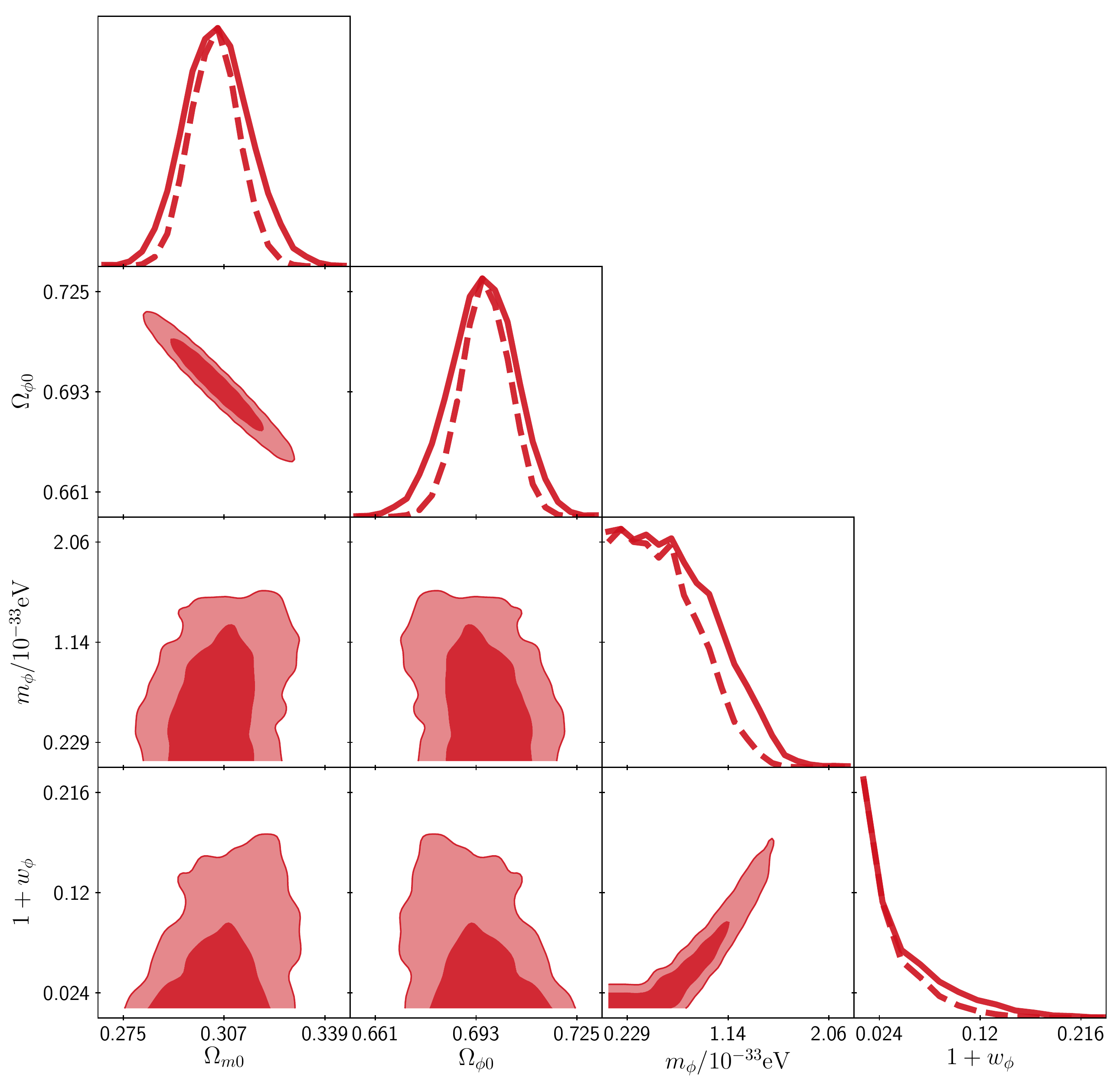}
\includegraphics[width=0.49\textwidth]{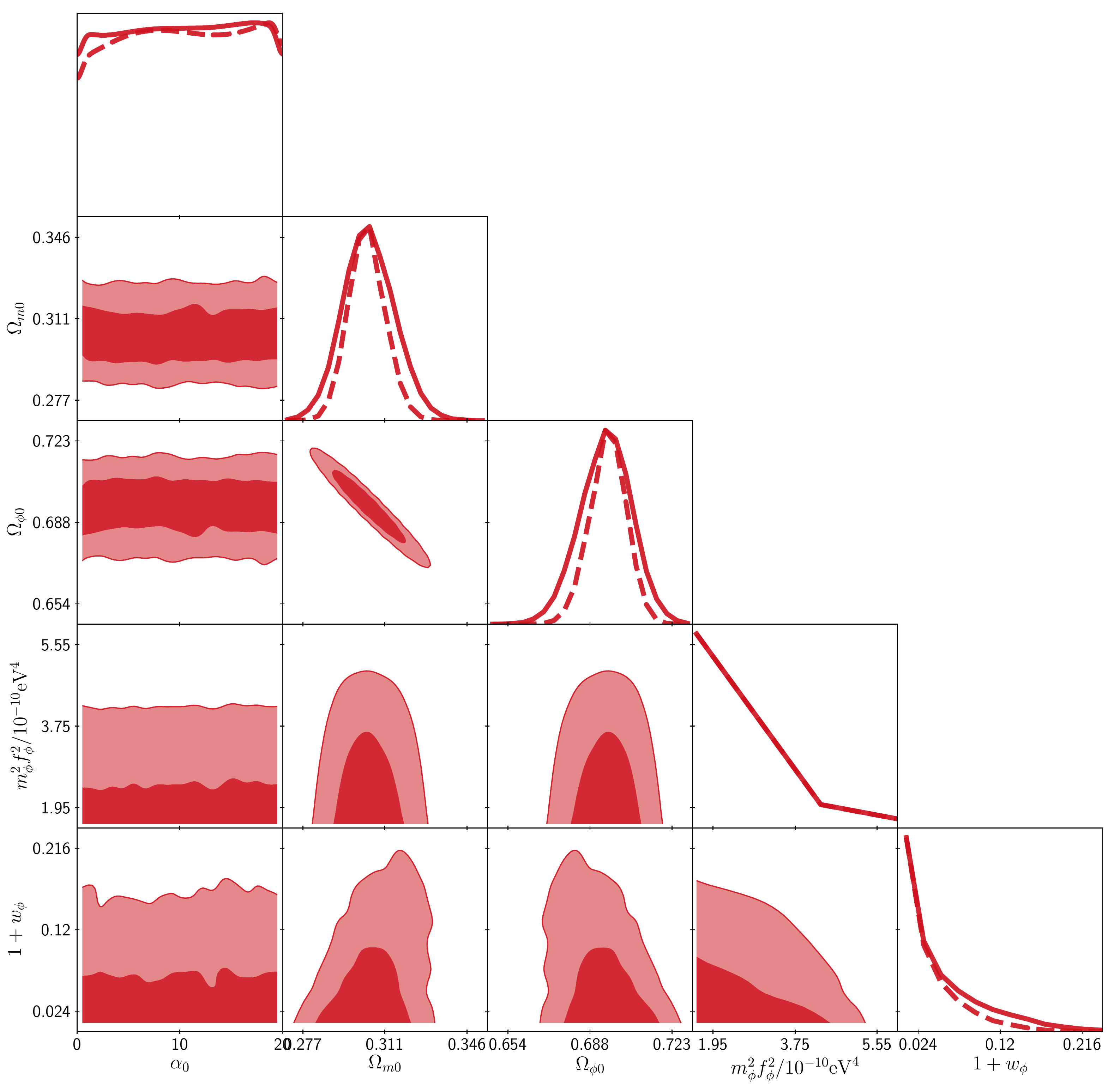}
\caption{\label{fig:triangle1} (Top) Posterior distributions (solid lines) and mean likelihood (dashed lines)  for  the constrained cosmological parameters corresponding to the quadratic potential in Class Ia. (Bottom) Same as above for the axion potential in Class IIa ($\alpha_1 = \alpha_2 =0$). As anticipated from Fig.~\ref{fig:triangle}, the dynamical variable $\alpha_0 = 3/(\kappa^2 f^2_\phi)$ remains unconstrained, whereas there appears an upper bound for the passive parameter $m^2_\phi f^2_\phi < 3.66 \times 10^{-10} \, \mathrm{eV}^4$. See the text for more details.}
\end{figure}

Taking as a reference the quadratic potential again, for which the mass of the scalar field $m_\phi$ is a passive parameter, we show in the top panel of Fig.~\ref{fig:triangle1} the posteriors of different cosmological quantities. The fit indicates that $\Omega_{m0} = 0.304^{+0.018}_{-0.016}$, $\Omega_{\phi 0} = 0.695^{+0.016}_{-0.018}$, $1+w_{\phi 0} < 0.129$, and $m_{\phi} < 1.36 \times 10^{-33} \, \mathrm{eV}$ (95\% C.L.). It can be seen that the preferred value of the EoS is close to $-1$, and that the scalar field mass $m_\phi$ has an upper bound. The latter constraint can be easily understood if we recall that the field mass can be calculated from the expression $m_\phi = (1/2) y_{10} H_0$, and then any bounds on the scalar field mass are directly obtained from those on the present values $y_{10}$ and $H_0$.

Other cases, though, are not as clear as the quadratic one. Let us consider the axion potential, which we write as $V(\phi) = m^2_\phi f^2_\phi \left[1 + \cos(\phi/f_\phi)\right]$, where $m_\phi$ is the mass of the axion field and $f_\phi$ is its so-called decay constant. The axion potential belongs to Class IIa with $\alpha_0 = 3/(\kappa^2 f^2_\phi)$  (together with $\alpha_1 = 0 = \alpha_2$), which indicates that, according to our classification, $f_\phi$ is a dynamical parameter, whereas the combination $m^2_\phi f^2_\phi$ forms a passive one. The immediate result is that $f_\phi$ cannot be constrained from cosmological observations, which is at odds with previous results in the literature (see \citep{Marsh:2015xka} and references therein). Our interpretation here is that previous studies were not able to sample all possible values of $f_\phi$, mostly because of the intrinsic difficulties in solving the quintessence field equations in their normal form, see Eq.~\eqref{eq:1d}. Smaller values of $f_\phi$ require the field to start closer to the top of the potential, and this is a tough numerical task even in our approach.

As for the passive parameter in the axion potential, it can be shown that it can be determined from\cite{Urena-Lopez:2015odd} 
\begin{equation}
m^2_\phi f^2_\phi = \frac{3H^2_0}{\kappa^2 \alpha_0} \left[ \frac{y^2_{10}}{4} + \alpha_0 \Omega_{\phi 0} \cos^2(\theta_0/2) \right] \, .
\end{equation}
The passive parameter of the quintessence potential cannot be written solely in terms of the cosmological observables ($H_0$) and the dynamical variables ($w_{\phi 0}$, $y_{10}$, $\rho_{\phi 0}$), and then it cannot be clearly constrained because of the presence of the decay constant $f_\phi$. Notice however that if $f_\phi \ll 1$ (in appropriate units), which corresponds to $\alpha_0 \to \infty$, then $m^2_\phi f^2_\phi \simeq \rho_{\phi 0} (1 - w_{\phi 0})/2$, where $\rho_{\phi 0}$ is the present quintessence density, and in this limit there appears an upper bound for the passive parameter basically inherited from the one on $\rho_{\phi 0}$. Likewise, if $f_\phi \gg 1$, corresponding to $\alpha_0 \to 0$, we find that $m_\phi f_\phi \simeq (y_{10} H_0/2) f_\phi$, which shows that the passive parameter in this limit will be unbounded from above and that $m_\phi \simeq (y_{10} H_0/2)$, which is exactly the result obtained for the quadratic potential.

The posterior distributions for the axion case are shown in the bottom panel of Fig.~\ref{fig:triangle1}. The fit indicates that $\Omega_{m0} = 0.288^{+0.037}_{-0.035}$, $\Omega_{\phi 0} = 0.712^{+0.035}_{-0.037}$, $1+w_{\phi 0} < 0.154$, and $m^2_{\phi} f^2_\phi < 3.66 \times 10^{-10} \, \mathrm{eV}^4$ (95\% C.L.). Apart from the upper bound for the passive parameter $m_\phi f_\phi$, the preferred values of the other parameters are similar to those of the quadratic potential (top panel in Fig.~\ref{fig:triangle1}) and also to those of the general case shown in Fig.~\ref{fig:triangle}. Hence, the study of particular cases does not provide stronger bounds for the cosmological parameters.

\section{Conclusions \label{sec:conclusions}}
In this work, we have presented a general study of quintessence dark energy models that allows a general comparison with observational data without the need to specify their functional form. This is possible because the equations of motion of the quintessence field are written as an autonomous system and later transformed to a polar form that automatically satisfies the Friedmann constraint. Moreover, one of the new dynamical variables in the polar form is directly related to the quintessence EoS, which then means that the latter is no longer a parameter derived from the field equations but rather one that controls the evolution of the quintessence field.

One interesting finding of this work is the general form of the quintessence potentials. To close the polar system of equations one needs the information about a second potential variable that we called $y_2$. The functional form of $y_2$ depends on the particular choice of quintessence potentials, but by observing the results obtained from different potentials we proposed a series form of $y_2$ that covers a wide range of models. We have correspondingly identified four different classes of quintessence potentials in terms of the series coefficients of $y_2$, which is integrated back to get the functional form of the quintessence potentials $V(\phi)$ that belong to the four classes.

We have found a general solution of the equation of motion in their polar form by taking into account the fact that the quintessence EOS is very close to $-1$ and subdominant in both the matter and radiation dominated eras. This solution is particularly interesting as it estimates the information about the initial conditions of the quintessence variables deep inside the radiation era by using the present values of the cosmological parameters. We have incorporated the expressions of the initial conditions in an amended version of the Boltzmann code CLASS, with which we have worked out the numerical solutions of the polar equations of motion. This has allowed us to find a parameterization for the evolution of the EoS that seems to suit better  the case of thawing quintessence than others proposed in the literature. The parametrization works well because is based on the analytical solutions found for the polar variables.

However, we did not consider the new parametrization of the EoS for a comparison with observations, but we rather worked directly with the polar equations of motion. According to our study of the quintessence potentials, we distinguished two separate set of parameters in them: the dynamical ones and the passive ones. The dynamical ones appear explicitly in the equations of motion and then have a direct influence on the evolution of the field variables. In contrast, the passive ones are integration constants that can be expressed as combinations of the polar variables and other cosmological variables like the Hubble parameter. The comparison with observations showed that the passive variables can in principle be constrained, but that is not the case of the dynamical parameters in the quintessence potentials, whose posteriors are fully flat. {We have verified that this is in agreement with other results already published in the literature}. This is one of our main results: that observations cannot establish a preference for a given functional form of the quintessence potential.

The dynamical variables were constrained but their allowed values are close to those of a cosmological constant and in this sense our analysis does not show any preference for quintessence models over a constant dark energy density. As a side result, we have also argued that the results on the dynamical variables can be used to put constraints on the passive parameters of the field potentials. This was done for a couple of particular examples, but our methods can be used for other types of potentials as well.

In all, our results indicate that there will be always a set of dynamical parameters which will satisfy the observational constraints for any given potential.  According to our method, this is because our current observations can only put an upper bound on the present value of the DE EoS,  $0 \leq 1+w_{\phi 0} < 0.107$ (in the general case, see Sec.~\ref{sec:general-results} above). The degeneracy in our results could be broken if there were any indication of a non-zero lower value in the EoS (which would, in turn, rule out a cosmological constant), as this will narrow the possible evolutionary paths of the quintessence variables and in consequence the allowed values of the dynamical variables. But given the current state of affairs, we cannot but to conclude that the problem with the arbitrariness of the functional form of the quintessence potential still remains unsolved.

\begin{acknowledgments}
N.R. acknowledges PRODEP for financial support. AXGM acknowledges support from C\'atedras CONACYT and UCMEXUS-CONACYT collaborative project funding. This work was partially supported by Programa para el Desarrollo Profesional Docente; Direcci\'on de Apoyo a la Investigaci\'on y al Posgrado, Universidad de Guanajuato, research Grants No. 206/2018; CONACyT M\'exico under Grants No. 286897, 182445, 179881, 269652 and Fronteras 281; and the Fundaci\'on Marcos Moshinsky. We also acknowledge the use of the computing facilities ATOCATL at UNAM, and COUGHS at UGTO. 
\end{acknowledgments}

\appendix
\section{General dynamical system approach for quintessence fields in terms of a roll parameter \label{sec:app-general}}
To show more about the convenience of Eq.~\eqref{GP} as a general representation of quintessence potentials, we first write
Eq.~(\ref{eq:4c}) as
\begin{equation}
\frac{y_1^{\prime}}{y} = \frac{3}{2} \left( 1 + w_{tot} \right)
\frac{y_1}{y} + x \frac{y_2}{y} \, , \label{eq:5}
\end{equation}
and in combination with Eqs.~\eqref{eq:4b} and~\eqref{GP} we find
\begin{eqnarray}
\left( \frac{y_1}{y} \right)^{\prime} &=& x \left( \frac{1}{2} \frac{y_1}{y} + \frac{y_2}{y} \right) 
\, . \label{eq:6}
\end{eqnarray}
If we use again the function defined in Sec.~\ref{sec:gener-form-quint}, $y_1/y = - (\sqrt{6}/\kappa) \partial_\phi \ln(V)
=  \lambda$, Eq.~\eqref{eq:6} can be written in the form\footnote{It must be noticed that $\lambda$ is related to the conventional roll parameter $\tilde{\lambda}$ in quintessence dynamical analysis as $\tilde{\lambda} = \lambda/\sqrt{6}$, where $\tilde{\lambda} = - (1/\kappa) \partial_\phi \ln(V)$, see for instance Refs.~\cite{Fang:2008fw,Chongchitnan:2007eb,UrenaLopez:2011ur,Copeland:1997et}.}
\begin{equation}
  \label{eq:9}
  \lambda^\prime = -x \lambda^2 \left[ \Gamma(\lambda) -1 \right] \, ,
\end{equation}
with $\Gamma \equiv V \partial^2_\phi V/(\partial_\phi V)^2$, which is
known as the tracking parameter\cite{Zlatev:1998tr,Scherrer:2007pu,Fang:2008fw,UrenaLopez:2011ur,Chongchitnan:2007eb}. A direct comparison between Eqs.~\eqref{eq:6} and~\eqref{eq:9} gives
\begin{equation}
   \frac{y_2}{y} = \lambda^2 \left[ 1 - \Gamma(\lambda) \right] - \frac{1}{2} \frac{y_1}{y} \, , \label{eq:15}
\end{equation}
which shows the direct relation between our new potential variable $y_2$ and the tracking parameter.

Some previous works have considered that for selected scalar field potentials there is a closed form of the tracking parameter $\Gamma(\lambda)$ in terms of $\lambda$\cite{Fang:2008fw}, and for those same potentials our dynamical system~\eqref{eq:4} becomes an autonomous one because $y_2 = y_2 (\theta,y_1,\Omega_\phi)$. Our method in this paper suggests that we may as well consider not the complete form but just a series expansion of $\Gamma(\lambda)$ to find general solutions of quintessence potentials.

Finally, Eqs.~\eqref{eq:4a} and~\eqref{eq:4b} are also rewritten as
\begin{subequations}
  \label{eq:2A}
  \begin{eqnarray}
    x^\prime &=& -3x + \frac{3}{2} \left(1 + w_{tot} \right) x + \frac{\lambda}{2} y^2 \, , \label{eq:2Aa} \\
    y^\prime &=&  \frac{3}{2} \left(1 + w_{tot} \right) y - \frac{\lambda}{2} x y \, , \label{eq:2Ab}
  \end{eqnarray}
\end{subequations}
which resemble the dynamical system of an exponential potential firstly studied in Ref.~\cite{Copeland:1997et}.

\section{Late time attractors \label{sec:app-late-time}}
Here we discuss about the late time attractor solutions of the dynamical system~\eqref{eq:4}. We are particularly interested in late time behaviour of the Universe hence we consider it to be dominated by dark matter and dark energy only. 

The fixed points of the systems can be find out by solving the three equations $\theta^\prime = 0$, $y_1 ^\prime = 0$, $\Omega^\prime = 0$ simultaneously. From the first of the conditions we find that at the critical point $y_{1c} = 3\sin \theta_c$. With this the equations of the critical points reduce to
\begin{subequations}
\label{eq:fixded}
  \begin{eqnarray}
\left[ 9 \left( 1  - \Omega_{\phi c} \cos \theta_c \right) + \Omega_{\phi c} \, (y_{2c}/y_c) \right] \sin\theta_c &=& 0 \, , \label{eq:fixed_b} \\ 
3 (1- \Omega_{\phi c}) \Omega_{\phi c} \cos \theta_c &=& 0 \, . \label{eq:fixed_c}       
  \end{eqnarray}
\end{subequations}

If we consider Eq.~\eqref{eq:fixed_c} we obtain either $\Omega_{\phi c} = 1$ (quintessence domination), $\Omega_{\phi c} =0$ (matter domination), or $\theta_c = \pi/2$. The latter solution is not unique, but for the purposes in this appendix we will restrict ourselves to the range $\theta = [0:\pi]$. We then need to solve Eq.~\eqref{eq:fixed_b} for the series expansion~\eqref{eq:10} to find all possible combinations of the critical values. The resultant values are summarized in Table~\ref{tab:5} for the series expansion of $y_2/y$ up to second order. We also indicate in the last column the classes of potentials from Table~\ref{tab:2} for which the given critical points can exist.

\begin{table*}[th!]
\caption{Fixed points and there corresponding eigenvalues. \label{tab:5}}
\squeezetable
\begin{ruledtabular}
\centering
\begin{tabular}{|c|c|c|c|c|}
Fixed points & $\Omega_{\phi c}$  & $\theta_c$ & $w_{\phi c}$ & Class \\
\hline 
$a$  & $0$  & $0,\pi$ & $-1,1$ & All Classes \\
 \hline
 
 $b$  & $1$  & $0,\pi$ & $-1,1$ & All Classes \\
 \hline
   
$c$ & $1 $& $\sin(\theta_c/2) = \frac{-\alpha_1 + \sqrt[]{\alpha_1 ^2 - 2 \alpha_0 (2 \alpha_2 +1)}}{6 (2 \alpha_2 +1)} $ & $2 \sin^2 (\theta_c/2) -1$ & Ia, IIa, IIIa, IVa \\

\hline
   
$d$ & $1 $& $ \sin(\theta_c/2) =  \frac{-\alpha_1 - \sqrt[]{\alpha_1^2 - 2 \alpha_0 (2 \alpha_2 +1)}}{6 (2 \alpha_2 +1)} $ & $2 \sin^2 (\theta_c/2) -1$  & Ia, IIa, IIIa, IVa \\

\hline
   
$e$ & $- \frac{3}{\sqrt[]{2} \alpha_0} \left[ \alpha_1 -  \sqrt[]{ \alpha_1^2 - \alpha_0 (2 \alpha_2 +1)} \right] $& $\pi/ 2$ & 0 & II, IV \\

\hline
   
$f$ & $- \frac{3}{\sqrt[]{2} \, \alpha_0} \left[ \alpha_1 + \sqrt[]{ \alpha_1^2 - \alpha_0 (2 \alpha_2 +1)}\right] $& $\pi/2$ & 0 & II, IV
\end{tabular}
\end{ruledtabular}
\end{table*}

\bibliography{quint}

 \end{document}